%% file: main.tex
\documentclass[journal]{IEEEtran}
\IEEEoverridecommandlockouts
\usepackage{cite}
\usepackage{amsmath,amssymb,amsfonts}
\usepackage{amsthm}
\usepackage{algorithm}
\usepackage{algpseudocode}

\usepackage{graphicx}
\graphicspath{{./figures/}}
\usepackage{textcomp}
\usepackage{xcolor}
\usepackage{booktabs}
\usepackage{multirow}
\usepackage{array}
\usepackage{soul}
\usepackage{tabularx}
\usepackage[table]{xcolor}
\usepackage{booktabs}
\usepackage{graphicx}
\usepackage{threeparttable}
\usepackage[normalem]{ulem}

\newcommand{\tool}{$\mathtt{Auriga}$}

\soulregister{\cite}{7}
\soulregister{\ref}{7}
\soulregister{\eqref}{7}
\soulregister{\tool}{0}

\usepackage[utf8]{inputenc}
\usepackage{listings}
\usepackage{xcolor}

\lstdefinelanguage{json}{
    basicstyle=\ttfamily\scriptsize,
    numbers=left,
    numberstyle=\tiny\color{gray},
    stepnumber=1,
    numbersep=4pt,
    showstringspaces=false,
    breaklines=true,
    frame=single,
    stringstyle=\color{red},
    keywordstyle=\color{blue}\bfseries,
}

\begin{document}

\title{\tool{}: Harness-Governed LLM Agents for Effective Throughput Optimization in Execute-Order-Validate Permissioned Blockchains}

\author{       
     Xiao Zhang, 
     Tianyang Wang,
     Fangtao Gu,
     Haoxiang Sun,
     Puwei Wang\textsuperscript{*}, 
     Jinchuan Chen\\ 
      
     {School of Information, Renmin University of China} \\
      
     \vspace{0.8mm}
     
     Email:\{zhangxiao, wangtianyang, fangtaogu, hxsun, wangpuwei, jcchen\}@ruc.edu.cn 
 }  
\maketitle

\input{sections/00_abstract}
\input{sections/01_introduction}

\input{sections/02_related_work}

\input{sections/03_overview}
\input{sections/04_tuning_agent}
\input{sections/05_scheduler}
\input{sections/06_evaluation}
\input{sections/08_conclusion}

\bibliographystyle{IEEEtran}
\bibliography{references}

\input{sections/09_author_biographies}

\end{document}

%% file: sections/00_abstract.tex
\begin{abstract}
Effective throughput, the product of raw throughput and transaction success rate, determines the useful capacity of Execute-Order-Validate blockchains such as Hyperledger Fabric. Most prior work either tunes configurations to raise raw throughput, requiring many costly redeployments, or controls conflicts to raise the success rate, repairing them only after execution with kernel modifications. The few joint approaches tune only a small set of parameters and still rely on kernel changes to resolve conflicts. To address these challenges, we present \tool{}, a non-invasive framework that jointly optimizes both factors through sample-efficient LLM-guided configuration search and pre-execution conflict prevention. A harness-governed multi-agent architecture coordinates LLM agents for parameter exploration, saturation-aware evaluation, and experience consolidation, with the harness controlling tool and memory access and retaining termination authority. For transaction scheduling, an LLM extracts symbolic AST templates from smart contracts offline. At runtime, the scheduler predicts read-write sets without executing them and partitions transactions into conflict-free groups. Across three benchmarks and conflict rates up to 70\%, \tool{} improves effective throughput by up to $6.0\times$ over default Fabric and $2.0\times$ over the strongest baseline. The tuning agent also discovers undocumented parameter dependencies, and ablation studies show the independent contributions of both subsystems.

\end{abstract}


%% file: sections/01_introduction.tex
\section{Introduction}
 
 Permissioned blockchains, particularly Hyperledger Fabric~\cite{fabric_arch}, have become the infrastructure of choice for enterprise deployments in finance, supply chain, and healthcare. They provide auditability and decentralization without the energy cost of proof-of-work consensus. Yet a barrier to production adoption is performance: under default configurations, Fabric networks routinely operate at a fraction of their hardware potential, and under realistic workloads with transaction contention, the \emph{effective throughput}, i.e., successful transactions committed per second, can degrade significantly~\cite{baliga_fabric,chacko_failures}.  

We observe that effective throughput is governed by two factors in a single equation:
\begin{equation}
    TPS_{eff} = TPS_{raw} \times \mathcal{R}_{succ} \label{eq:efftps}
\end{equation}

Raw throughput ($TPS_{raw}$) is governed by hundreds of interacting parameters across block cutting, concurrency limits, and gossip propagation~\cite{fik}. Their optimal values depend on hardware and network topology, leaving default deployments with substantial underutilized capacity~\cite{athena}.

Conversely, the transaction success rate ($\mathcal{R}_{succ}$) is limited by Multi-Version Concurrency Control (MVCC) conflicts inherent to the Execute-Order-Validate (EOV) architecture~\cite{chacko_failures,debreczeni_taxonomy}. When concurrent transactions within the same block conflict over identical keys, only one commits, forcing the rest to abort after wasting expensive endorsement and consensus pipeline resources~\cite{fabricsharp}. 

The two performance bottlenecks compound under intensive workloads, as quantified in Tab.~\ref{tab:motivation_table}. At a 40\% conflict rate, state-of-the-art configuration tuning~\cite{athena} lifts raw infrastructure capacity to improve effective throughput from 205 to 612. However, because it lacks concurrency-awareness, its success rate remains at 67.8\%, wasting over 32.2\% of computing and consensus resources on aborted transactions. In contrast, by optimizing static configurations and proactive scheduling, \tool{} maintains a 90.2\% success rate under the identical conflict rate, achieving an effective throughput of 1{,}226 to deliver a $6.0\times$ acceleration over the default deployment.
  
\begin{table}[htbp]
\caption{Motivation: Performance under Varying Contention}
\label{tab:motivation_table}
\centering
\resizebox{\columnwidth}{!}{%
\begin{tabular}{lcccc}
\toprule
\textbf{Framework} & \multicolumn{2}{c}{\textbf{10\% Conflict}} & \multicolumn{2}{c}{\textbf{40\% Conflict}} \\
\cmidrule(r){2-3} \cmidrule(l){4-5}
\textbf{Configuration} & $\mathcal{R}_{succ}$ & $TPS_{eff}$ & $\mathcal{R}_{succ}$ & $TPS_{eff}$ \\
\midrule
Default Fabric Config & 91.4\% & 294 & 63.1\% & 205 \\
Configuration Tuning Only \cite{athena} & 88.9\% & 1,061 & 67.8\% & 612 \\
\rowcolor[gray]{0.9} \tool{}  & \textbf{97.7\%} & \textbf{1,348} & \textbf{90.2\%} & \textbf{1,226} \\
\bottomrule
\end{tabular}%
}
\end{table}

In automated parameter configuration, methodologies have advanced from heuristic sampling and Bayesian optimization to deep reinforcement learning (DRL) and large language model (LLM) agents~\cite{bestconfig,ottertune,cdbtune,gptuner,agenttune}. However, these approaches suffer from an evaluation overhead mismatch in permissioned blockchains. Unlike traditional databases that support near-instantaneous parameter updates, evaluating a single configuration set in a blockchain requires updating parameters across all nodes and restarting the entire network. This expensive process makes sample-heavy DRL or exhaustive tree search impractical. Furthermore, existing LLM-assisted tuning systems like AgentTune~\cite{agenttune} are primarily tailored for database engines. Without rigid, harness-governed behavioral guards, adapting these frameworks to distributed environments can introduce operational risks or slow convergence. BCTuner~\cite{deng2026bctuner} grounds an LLM in system manuals and hardware specifications to guide Monte Carlo tree search over blockchain configurations. However, each tuning action adjusts only one cluster of related knobs or a single knob, rather than revising coupled parameters across clusters at once. It also evaluates candidates at a fixed send rate, which may not reflect a configuration's actual capacity.

Conversely, conflict-control approaches focus on mitigating MVCC aborts to improve transaction success rates~\cite{fabricpp,fabricsharp,fabrictl}. Though effective in specific settings, these techniques exhibit two primary limitations. First, they often introduce intrusive modifications to ordering mechanisms or peer validation pipelines~\cite{fabricsharp}, constraining cross-version compatibility and complicating deployment. Second, their conflict mitigation is typically reactive or execution-dependent~\cite{xoxfabric,song2024htfabric}. Because they rely on parsing read-write sets from prior transaction simulations, they are less equipped to preemptively manage data contention before the execution phase begins.

\textbf{Central gap.}
Prior work within the EOV stack optimizes one factor of Eq.~\eqref{eq:efftps} at a time. Tuning approaches~\cite{fik,athena,deng2026bctuner} lift raw throughput but assume conflict-free workloads, overlooking the success-rate multiplier; conflict-control approaches~\cite{fabricpp,fabricsharp,fabrictl,xoxfabric,song2024htfabric} raise the success rate but assume pre-optimized configurations, repair conflicts only after execution, and require intrusive kernel modification. TuneChain~\cite{tunechain} jointly addresses both factors, yet the techniques it combines, namely sample-hungry reinforcement learning and post-execution conflict reordering, limit what it can achieve. Its RL tuner is confined to a few restart-free knobs, leaving the parameter space that governs raw capacity unexplored; its read-only-first scheme reorders transactions by type without identifying which transactions actually conflict, so its benefit shrinks as conflicts concentrate among write transactions, and aborted transactions still waste pipeline resources. Both components further modify the blockchain kernel. In contrast, \tool{} jointly optimizes both factors: an LLM-guided, sample-efficient search explores a broad range of the parameter space offline, and a pre-execution scheduler prevents conflicts before they occur, all without modifying the blockchain kernel.

Orthogonal to in-stack optimization, architecture/protocol adaptation leaves the EOV stack altogether: AdaChain~\cite{adachain} switches among candidate architectures at runtime, and BFTBrain~\cite{bftbrain} switches among BFT consensus protocols. We set this direction aside because production Fabric deployments carry locked-in ledgers, smart contracts, and compliance obligations that make stack replacement infeasible, leaving in-stack optimization as the practical problem (see Sec.~\ref{sec:related}).

Realizing this unified framework introduces two primary challenges. First, optimizing hundreds of static parameters triggers a combinatorial explosion, rendering conventional trial-and-error benchmarks unviable due to expensive network restarts. Second, achieving MVCC awareness requires predicting each transaction's read-write set before execution, without making intrusive modifications to the blockchain kernel. To address these challenges, our central insight is that static configurations define baseline capacity while dynamic scheduling prevents conflict-induced waste, requiring their synthesis into a unified framework.
 
To realize this synergy, we propose \tool{}, a plug-and-play framework that jointly optimizes the two factors of Eq.~\eqref{eq:efftps} through configuration tuning and transaction scheduling. \tool{} overcomes the aforementioned challenges via a split-phase pipeline. In the static tuning phase, it deploys a harness-governed multi-agent architecture coordinated through a shared runtime. Grounding the LLM agents in a knowledge base of parameter semantics and coupling dependencies, the framework prunes hazardous actions to efficiently identify optimal configurations. In the dynamic scheduling phase, \tool{} proactively predicts each transaction's read-write set using offline-generated AST (Abstract Syntax Tree) templates and world-state statistics. It then constructs a conflict graph among transactions and groups non-conflicting ones into batches before submission, avoiding MVCC aborts without modifying the blockchain kernel. The batches are injected at the calibrated send rate.

We evaluate \tool{} across diverse benchmarks against multiple state-of-the-art baselines. Experimental results show that \tool{} delivers up to a $6.0\times$ effective throughput improvement over the default Fabric configuration and up to a $2.0\times$ acceleration over the best-performing standalone baseline in each scenario. Ablation studies show that both the configuration tuning and transaction scheduling subsystems independently outperform specialized baselines in their respective subtasks. Notably, the LLM agent uncovers a critical dependency between two gossip-layer parameters (Sec.~\ref{sec:eval_tuning}), overlooked by prior literature.

In summary, the contributions of this paper are as follows:
\begin{itemize}    

    \item \textbf{Harness-Governed Multi-Agent Configuration Tuning}: An LLM agent-based framework that separates configuration exploration, saturation-aware evaluation, and experience consolidation into three specialized agents, shielding the reasoning context from raw runtime logs while enabling structured knowledge reuse across trials.

    \item \textbf{LLM-Driven Predictive Transaction Scheduling}: Offline LLM-based extraction of symbolic AST templates and required world-state statistics; online prediction of read-write sets and conflict-free batching, avoiding MVCC aborts without modifying the blockchain core.

    \item \textbf{System Implementation and Evaluation}: We implement \tool{}, integrating configuration tuning and transaction scheduling. Experiments across diverse topologies, workloads, and conflict rates consistently show that holistic co-optimization outperforms standalone and hybrid baselines, supporting the necessity and effectiveness of the proposed approach.
\end{itemize}
 
The remainder of the paper is organized as follows.

 Sec.~\ref{sec:related} surveys related work and positions \tool{} along three axes, as summarized in Table~\ref{tab:related-work}. Sec.~\ref{sec:overview} presents the system architecture. Sec.~\ref{sec:tuning} and~\ref{sec:scheduler} detail the Configuration Tuning and Transaction Scheduling subsystems, respectively. Sec.~\ref{sec:evaluation} reports our experimental evaluation, and Sec.~\ref{sec:conclusion} concludes.

%% file: sections/02_related_work.tex
\section{Related Work}
\label{sec:related}

\begin{table*}[!t]
\centering
\renewcommand{\arraystretch}{1.1}
\footnotesize
\begin{threeparttable}
\caption{Positioning of Auriga among Performance-Optimization Approaches for Permissioned Blockchains}
\label{tab:related-work}
\begin{tabularx}{\textwidth}{@{}l X c c c@{}}
\toprule
\textbf{System} & \textbf{Method} & \shortstack{\textbf{Config.}\\\textbf{Tuning}} & \shortstack{\textbf{Conflict}\\\textbf{Control}} & \shortstack{\textbf{Kernel}\\\textbf{Modification}} \\
\midrule
\multicolumn{5}{@{}l}{\textbf{Configuration Tuning}} \\[2pt]
FIK~\cite{fik} & Sampling-based; Latin hypercube sampling with importance ranking of knobs & \checkmark & $\times$ & No \\
Athena~\cite{athena} & Reinforcement learning; multi-agent DDPG over peer and orderer knobs & \checkmark & $\times$ & No \\
Zhai et al.~\cite{zhai_explainable_drl} & Reinforcement learning; causal-model-enhanced DRL for explainable tuning & \checkmark & $\times$ & No \\
BCTuner~\cite{deng2026bctuner} & LLM; LLM-guided Monte Carlo tree search over configurations & \checkmark & $\times$ & No \\
\addlinespace
\multicolumn{5}{@{}l}{\textbf{Conflict Control}} \\[2pt]
Fabric++~\cite{fabricpp} & Post-execution; intra-block reordering with early abort & $\times$ & \checkmark & Yes \\
FabricSharp~\cite{fabricsharp} & Post-execution; dependency-graph filtering of unserializable transactions before ordering & $\times$ & \checkmark & Yes \\
FabricTL~\cite{fabrictl} & Post-execution; topological sorting weighted by transaction influence & $\times$ & \checkmark & Yes \\
XOX Fabric~\cite{xoxfabric} & Post-execution; adds a re-execution phase after ordering for conflicting transactions & $\times$ & \checkmark & Yes \\
HTFabric~\cite{song2024htfabric} & Post-execution; fast reordering with parallel re-execution & $\times$ & \checkmark & Yes \\
\addlinespace
\multicolumn{5}{@{}l}{\textbf{Joint Optimization of Effective Throughput}} \\[2pt]
TuneChain~\cite{tunechain} & Reinforcement learning; multi-instance PPO with causal-model action filtering; post-execution, read-only-first block reordering toggled as a tuning knob & \checkmark & \checkmark & Yes \\
\textbf{Auriga (ours)} & \textbf{LLM; harness-governed multi-agent search with saturation-aware evaluation; pre-execution, AST-template-based read-write-set prediction and conflict-free grouping before submission} & \textbf{\checkmark} & \textbf{\checkmark} & \textbf{No} \\

\bottomrule
\end{tabularx}
\begin{tablenotes}
\footnotesize
\item[] \checkmark: addressed; $\times$: not addressed. \textbf{Kernel Modification}: whether it modifies the EOV ordering/validation pipeline (Yes) or operates outside it (No).
\item[] \textbf{Scope}: the three categories cover the approaches within the EOV stack; alternative architectures and system-restructuring designs lie outside this scope.
\end{tablenotes}
\end{threeparttable}
\end{table*}

We scope this paper to performance optimization of permissioned blockchains built on the Execute-Order-Validate (EOV) architecture. In such chains, transactions are first executed against the current state to produce read-write sets, then ordered by a single consensus service, and finally validated against those read-write sets, which aborts conflicting transactions. We consider only approaches that keep the EOV pipeline intact and optimize performance within it, rather than replacing the architecture. These fall into three categories, summarized in Table~\ref{tab:related-work}: configuration tuning, which raises the raw-capacity factor of Eq.~\eqref{eq:efftps}; conflict control, which raises the success-rate factor; and joint optimization of both. We cover these three categories. For configuration tuning, we additionally trace its lineage in database tuning; these works inform our design but are not entries in the table.

Beyond EOV, two alternative architectures place execution at different points relative to ordering~\cite{fabric_arch}. Order-Execute (OE) chains order transactions first and execute them deterministically afterwards, while Execute-Order-Execute (EOE) chains re-execute every transaction after ordering in place of validation, eliminating read-write conflicts by construction. A further line, architecture/protocol adaptation, treats the pipeline itself as the tuning lever: AdaChain~\cite{adachain} switches at runtime among candidate architectures that include EOV-based Fabric deployments, and BFTBrain~\cite{bftbrain} switches among BFT consensus protocols. All of these systems still run a single consensus instance at a time; they differ only in where execution sits relative to ordering or in which pipeline is selected. Approaches farther afield (sharding, which partitions state and consensus among parallel committees~\cite{luu2016secure,wang2019monoxide}; DAG-based designs, which decouple data dissemination from the ordering instance~\cite{danezis2022narwhal,spiegelman2022bullshark}; and Layer-2 protocols, which move execution into off-chain channels and rollups~\cite{poon2016bitcoin,kalodner2018arbitrum}) restructure the system rather than optimize within a single pipeline and are therefore orthogonal to our scope.

\subsection{Configuration Tuning}

\subsubsection{Traditional Database Parameter Tuning}
Parameter tuning for data-intensive systems has evolved across three major paradigms: heuristic sampling~\cite{bestconfig}, Bayesian optimization (BO)~\cite{ottertune}, and deep reinforcement learning (DRL)~\cite{cdbtune}. 

\subsubsection{LLM-Enhanced System Tuning}
LLM-based approaches have entered system optimization, though existing frameworks target database engines. One paradigm, exemplified by GPTuner~\cite{gptuner}, treats LLMs as passive knowledge extractors to constrain Bayesian optimization spaces. Another paradigm adopts a one-shot configuration approach, where $\lambda$-Tune~\cite{lamdaTune} and E2ETune~\cite{e2etune} generate configuration files in a single inference call to trade iterative search for speed. Recently, AgentTune~\cite{agenttune} iteratively recommends knobs via a multi-agent workflow with tree-based beam search.

Both lineages, however, presuppose cheap trial evaluations, which do not hold in permissioned blockchains: each configuration trial requires updating parameters across all nodes and restarting the entire network. This cost motivates \tool{}'s knowledge-guided, sample-efficient search.

\subsubsection{Tuning for Permissioned Blockchains}
Within the blockchain domain, tuning frameworks in the Configuration Tuning category of Table~\ref{tab:related-work} have progressed from empirical micro-benchmarks~\cite{baliga_fabric} and queuing-network models~\cite{wu_queuing} to automated optimization. FIK~\cite{fik} identifies performance-critical parameters via Latin hypercube sampling. Athena~\cite{athena} applies Multi-Agent DDPG to tune peer and orderer configurations, while Zhai~\textit{et al.}~\cite{zhai_explainable_drl} integrate a causal model into DRL for explainability. BCTuner~\cite{deng2026bctuner} grounds an LLM in system manuals and hardware specifications to guide Monte Carlo tree search over configurations, in which each tuning action adjusts either one cluster of related knobs or a single knob. \tool{} imposes no such restriction: each hypothesis may jointly adjust any set of parameters across system layers, and the LLM relates evidence across trials to decide which to change. BCTuner also evaluates candidates at a fixed send rate rather than driving each one to saturation. Whether sampling-, DRL-, or LLM-guided, these methods optimize raw capacity alone: none of them addresses the success-rate factor of Eq.~\eqref{eq:efftps}, leaving MVCC-induced resource waste unmanaged.

\subsection{Conflict Control}

\subsubsection{Post-Execution Conflict Repair}
The five systems in the Conflict Control category of Table~\ref{tab:related-work} operate strictly post-execution: read-write sets are produced by simulation first, and conflicts are repaired only afterwards. Three of the five repair conflicts by reordering transactions within a block. Fabric++~\cite{fabricpp} pioneers this direction by adapting database concurrency control to blockchains, implementing intra-block transaction reordering and early transaction aborts to deliver throughput gains. FabricSharp~\cite{fabricsharp} extends this paradigm by applying optimistic concurrency control principles, filtering unserializable transactions prior to the ordering phase to minimize conflicts. FabricTL~\cite{fabrictl} accelerates the sequencing computation itself, introducing topological sorting weighted by transaction influence to reduce sorting-phase latency compared to Fabric++. Rather than reordering, XOX Fabric~\cite{xoxfabric} adds a post-ordering re-execution phase for conflicting transactions. HTFabric~\cite{song2024htfabric} combines both mechanisms, pairing fast reordering with parallel re-execution.

These five systems nevertheless diverge from \tool{} along the two dimensions recorded in Table~\ref{tab:related-work}. First, being post-execution, they are inherently reactive: conflicts can only be repaired after read-write sets have been observed, and transactions whose conflicts cannot be repaired in-block still abort. \tool{} instead predicts read-write sets prior to execution via AST-template synthesis and separates conflicting transactions into conflict-free groups before submission, preventing the failures these systems merely repair. Second, all five modify the blockchain kernel (e.g., ordering/validation pipelines), tying gains to specific platform versions, whereas \tool{} runs entirely with no kernel changes.

\subsection{Joint Optimization of Effective Throughput}
TuneChain~\cite{tunechain}, a representative work in the Joint Optimization category of Table~\ref{tab:related-work}, is the closest prior work to \tool{}: it models \textit{effective throughput} and jointly addresses both factors of Eq.~\eqref{eq:efftps}, coupling multi-instance PPO tuning with a conflict emergency mechanism that prioritizes read-only transactions over write transactions within a block, exposed as a binary tuning knob. It nevertheless differs from \tool{} in three essential ways. First, the mechanism is reactive and type-level: it never identifies which transactions actually conflict, so its benefit vanishes when conflicts arise among write transactions, the dominant pattern in transfer-style and batch-write workloads, and read-modify-write contention on the same key cannot be resolved by intra-block reordering at all. Second, TuneChain confines its action space to restart-free knobs (block size, block interval, and the reordering switch), leaving the broader parameter space that governs raw capacity, including gossip, communication, and execution settings, unexplored, whereas \tool{} explores these parameters offline while its scheduler adapts online. Third, the mechanism modifies the ordering service, whereas \tool{} leaves the blockchain kernel unmodified.

In summary, \tool{} stands apart from the systems surveyed above. Where the conflict-control systems repair conflicts only after execution, \tool{} prevents them before execution; and where TuneChain jointly addresses both factors of Eq.~\eqref{eq:efftps} but modifies the blockchain kernel, \tool{} does so while operating outside it. It introduces proactive AST-template synthesis and partitioning to resolve MVCC conflicts prior to execution, and combines harness-governed multi-agent configuration search with predictive scheduling, thereby improving effective throughput across deployment environments.

%% file: sections/03_overview.tex
\section{\tool{} Overview}
\label{sec:overview}

\begin{figure}[ht]
\centerline{\includegraphics[width=1\linewidth]{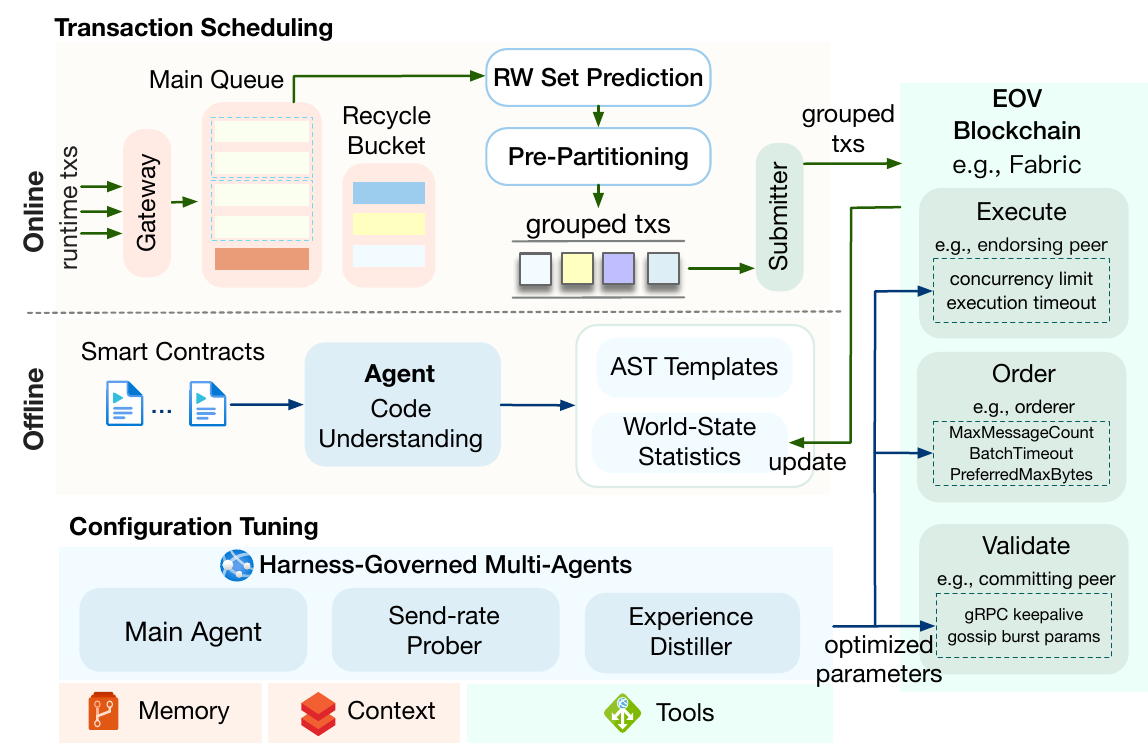}}
\caption{Overview of \tool{}}
\label{fig:overall}
\end{figure}

\tool{} is designed to maximize \emph{effective throughput} in EOV blockchains without modifying the kernel mechanisms. As defined in Eq.~\eqref{eq:efftps}, effective throughput is the product of two factors: the raw processing throughput $TPS_{raw}$ and the MVCC validation success rate $\mathcal{R}_{succ}$. \tool{} therefore combines two complementary optimizations, as shown in Fig.~\ref{fig:overall}: \emph{Configuration Tuning} lifts the $TPS_{raw}$ factor by expanding the achievable transaction processing capacity, while \emph{Transaction Scheduling} lifts the $\mathcal{R}_{succ}$ factor by reducing intra-block MVCC conflicts.

To lift both factors, \tool{} organizes its optimization workflow into offline preparation and online execution while remaining \emph{external} to the EOV blockchain. In the offline stage, \emph{Configuration Tuning} explores the configuration space to determine a resource-efficient operating point, characterized by the transaction arrival rate $\mathtt{send\_rate}$ and the maximum transaction count per block {\small\texttt{MaxMessageCount}}. Meanwhile, \emph{Transaction Scheduling} analyzes smart contracts to extract symbolic AST templates and identify the world-state statistics required for read-write-set prediction. During online execution, \emph{Transaction Scheduling} receives runtime transactions through its \emph{Gateway}, predicts their read-write sets using the offline-generated AST templates, and organizes them into conflict-free groups. The resulting groups are passed to the \emph{Submitter}, which forwards them to the underlying EOV blockchain. Thus, the computationally intensive configuration exploration and code analysis are performed offline, while only lightweight prediction and scheduling remain on the transaction path.

\subsection{Configuration Tuning}

Configuration tuning operates offline and aims to maximize the sustainable transaction processing rate. 
In practice, EOV blockchains expose hundreds of tunable parameters spanning concurrency limits, execution timeouts, block-cutting thresholds such as {\small\texttt{MaxMessageCount}}, and gRPC keepalive settings~\cite{fik}. These parameters are highly coupled across system layers, so the configuration space is difficult to navigate.

LLM agents can coordinate exploration and decision-making for complex optimization tasks~\cite{coala,swe_agent}. However, configuration tuning in EOV blockchains introduces two practical challenges: a configuration's merit is revealed only near its throughput limit rather than at an arbitrary arrival rate, and each trial requires costly redeployment followed by sustained measurement. Consequently, optimization must extract maximum information from a limited number of trials while preserving knowledge from previous evaluations.

To address these challenges, \tool{} adopts a multi-agent architecture that separates configuration exploration, throughput probing, and experience consolidation (Fig.~\ref{fig:overall}). The \emph{Main Agent} explores candidate configurations and coordinates the tuning process. The \emph{Send-rate Prober} evaluates configurations under increasing transaction arrival rates to identify their sustainable operating points. The \emph{Experience Distiller} consolidates observations from completed trials into reusable knowledge that guides subsequent exploration. These three agents are orchestrated by a dedicated \emph{Harness} runtime that mediates their tool and memory access and retains sole termination authority. This keeps autonomous LLM exploration within safe and reproducible bounds. The tuning process ultimately determines the operating point for the online scheduler. Since tuning acts purely through system settings, the underlying EOV implementation remains unchanged.

The detailed design is presented in Section~\ref{sec:tuning}.

\subsection{Transaction Scheduling}

Transaction scheduling aims to reduce intra-block MVCC conflicts by \emph{proactively} organizing transactions before block assembly. In EOV blockchains, conflicts are identified only during validation, after execution and ordering have already consumed system resources; conflicting transactions within the same block therefore commit nothing despite these costs.

\tool{} addresses this inefficiency by predicting transaction read-write sets before block assembly and organizing transactions by predicted conflict potential. This prediction, however, faces an information asymmetry at the gateway. A transaction's read-write set depends on which execution path it takes; the path taken depends on the transaction's arguments and the current world state, which is unknown at submission time.

The scheduling subsystem consists of an offline analysis stage and an online gateway-side scheduling stage. In the offline stage, an LLM-based agent analyzes smart contracts to identify the execution paths that affect ledger accesses, and from them constructs symbolic AST templates and specifies the world-state statistics required for runtime estimation.

In the online stage, runtime transactions first enter the scheduler through the \emph{Gateway} and are buffered in the main queue. The scheduler instantiates the offline-generated AST templates with transaction arguments and cached world-state statistics, then predicts their read-write sets. Based on the predicted access patterns, it performs \emph{Pre-Partitioning} to organize transactions into groups that are conflict-free with respect to the predicted read-write sets, each bounded by the tuned block capacity ({\small\texttt{MaxMessageCount}}). A \emph{Recycle Bucket} buffers transactions that fit no compatible group for reassembly into subsequent ones. The resulting transaction groups are forwarded to the \emph{Submitter}, which submits each group as a block to the underlying EOV blockchain through its standard transaction interface. The scheduler therefore operates entirely outside the blockchain core; the endorsing, ordering, and committing peers remain unmodified.

The detailed scheduling mechanism is presented in Section~\ref{sec:scheduler}.
         

%% file: sections/04_tuning_agent.tex
\section{Harness-governed Configuration Tuning}
\label{sec:tuning}

\begin{figure}[ht]
\centerline{\includegraphics[width=1\linewidth]{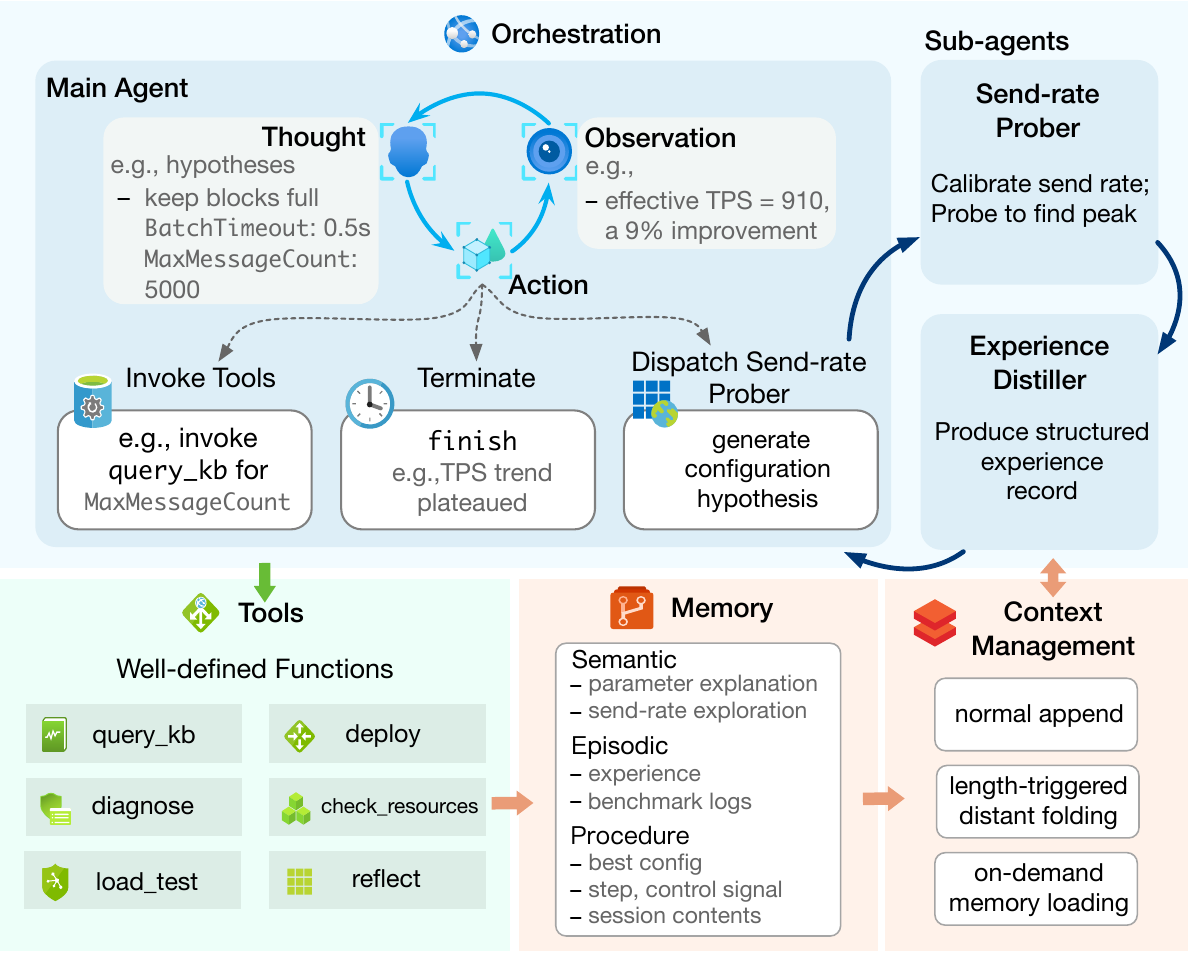}}
\caption{Harness-governed Configuration Tuning.}
\label{fig:tuning}
\end{figure}

The configuration tuning challenges necessitate a closed-loop process that combines throughput-aware evaluation with long-horizon experience accumulation, two capabilities that operate at different temporal scales. Existing LLM-based tuners either let a single agent free-run, exposing the deployment to unsafe actions and premature termination~\cite{agenttune}, or hard-code the search procedure so that the LLM merely answers stage-specific prompts~\cite{deng2026bctuner}. As shown in Fig.~\ref{fig:tuning}, \tool{} employs a harness-governed multi-agent architecture that assigns configuration exploration, saturation-aware evaluation, and experience consolidation to three specialized agents orchestrated by a shared runtime. The Harness structures the agent interface and information flow, while the LLM steers the search itself. This decomposition confines each agent to a well-defined decision scope and enables reuse of structured knowledge across costly trials.

\subsection{Multi-Agent Design}

Following the role-separation principle of multi-agent frameworks~\cite{metagpt}, \tool{} decomposes the tuning workload across three specialized agents, each detailed below.

\subsubsection{Main Agent}
\label{sec:main_agent}

The agent navigates the configuration space through a ReAct cycle~\cite{react}. 

\textbf{Thought.}
The agent reasons over its accumulated context and may form one or more of the following reasoning intents. When available information is insufficient, it identifies a knowledge gap and \emph{requests additional knowledge}. When current evidence suggests a promising direction, it formulates a \emph{configuration hypothesis} for the next step. When evaluation metrics fall short or degrade, it performs \emph{performance diagnosis} to isolate the underlying bottlenecks. When exploration yields diminishing returns or the search stabilizes, it \emph{assesses stagnation and convergence}, reconsidering its direction and whether further exploration is warranted. These intents condition the subsequent action but do not map one-to-one to it.

\textbf{Action.}
The agent executes the intent formed in the preceding thought, selecting from three
action types:

\begin{itemize}
\item \emph{Dispatch Send-rate Prober}.
The agent delegates a configuration hypothesis to the Prober for saturation-aware evaluation, the primary exploration action. 

\item \emph{Invoke tools}.
The agent invokes its tools: $\mathtt{query\_kb}$ retrieves parameter semantics and coupling dependencies; $\mathtt{diagnose}$ summarizes compressed runtime logs; $\mathtt{check\_resources}$ collects hardware utilization during load tests; and $\mathtt{reflect}$ reloads the experience store to reveal overlooked parameter regions.

\item \emph{Terminate}.
When convergence is deemed sufficient, the agent terminates the search.
\end{itemize}

\textbf{Observation.}
Each action yields an observation that updates the agent's context and conditions the next thought:
\begin{itemize}
    \item \emph{After configuration evaluation}, the observation comprises the result returned by the \emph{Send-rate Prober}: the measured peak sustainable throughput, or a failure report with compressed logs for subsequent diagnosis.
    \item \emph{After tool invocation}, the observation is the tool's output: structured parameter knowledge from $\mathtt{query\_kb}$, a diagnostic summary from $\mathtt{diagnose}$, hardware-utilization readings from $\mathtt{check\_resources}$, or a historical trajectory synthesis from $\mathtt{reflect}$.
\end{itemize}

Conditioned on the message history, the cycle progressively steers the search toward high-performing configurations.

\subsubsection{Send-rate Prober}
\label{sec:sa1}

The \emph{Send-rate Prober} evaluates a configuration hypothesis under saturation conditions. For a given configuration, raw throughput increases with the offered $\mathtt{send\_rate}$ until a topology-dependent ceiling, beyond which the network overloads and performance collapses sharply. A configuration tested below its saturation point therefore appears suboptimal, while one tested above may already have degraded. This saturation-aware protocol contrasts with fixed-rate evaluation~\cite{deng2026bctuner}: it makes configuration rankings invariant to the tester's choice of workload intensity (cf. Sec.~\ref{sec:eval_tuning}).

To obtain the throughput ceiling, the Prober executes a two-phase calibration for each hypothesis.

\begin{itemize}
    \item \emph{Deployment.} The Prober invokes the $\mathtt{deploy}$ tool to deploy the blockchain network with the proposed configuration and then the $\mathtt{load\_test}$ tool to run an initial benchmark at a conservative $\mathtt{send\_rate}$. This run verifies the configuration and establishes a throughput reference for subsequent probing.

    \item \emph{Saturation probing.} The Prober probes the deployed network by repeatedly invoking the $\mathtt{load\_test}$ tool at varying $\mathtt{send\_rate}$. It first uses coarse steps to bracket the saturation region, then switches to finer steps within that region to locate the peak sustainable throughput. Step sizes are adjusted to each topology: where throughput drops sharply beyond saturation, finer steps keep the Prober from overshooting the peak and underestimating throughput.

    \end{itemize}

\subsubsection{Experience Distiller}
\label{sec:sa2}

After each configuration evaluation, the \emph{Experience Distiller} produces a structured  record for the completed iteration. The Distiller receives the previous and current (configuration, TPS) pairs and generates a delta record capturing: which parameters changed, the direction and magnitude of each change, the observed TPS delta, and a causal hypothesis. This append-only experience store serves two purposes. First, it provides a compact summary of every trial that persists independently of the \emph{Main Agent}'s context window, so distant turns can be compressed as iterations accumulate. Second, by documenting configurations that proved inferior, it creates a searchable record that keeps the agent from revisiting unproductive regions of the parameter space. 

\subsection{Harness Runtime}

The Harness Runtime is the execution environment that hosts and orchestrates the three agents. It enforces the boundaries between their roles, preserves state across trials, and drives the whole lifecycle through three integrated mechanisms: orchestration, tool mediation, and memory and context management.

\subsubsection{Orchestration}

The Harness presents the accumulated message history to the \emph{Main Agent}, which proposes a configuration hypothesis, a tool call, or a termination request. A hypothesis is forwarded to the \emph{Send-rate Prober}, whose saturation-aware evaluation returns a compact summary of the peak throughput; the Harness then invokes the \emph{Experience Distiller} to produce a structured delta record and returns control to the \emph{Main Agent}. Tool calls are executed by the Harness directly and returned immediately. 

This mediation provides two properties essential for reliable long-horizon search. First, the Harness is the authority on termination: the \emph{Main Agent} may request termination, but only the Harness can grant it after verifying exploration depth, group diversification, and trend stabilization, thereby avoiding premature convergence from a single agent's judgment~\cite{agenttune}. Second, every trial outcome is recorded before the loop advances, so no evidence is lost to crashes or context overflow. Once termination is granted, the Harness emits the final configuration and its probed sustainable $\mathtt{send\_rate}$, which parameterizes the online scheduler (Section~\ref{sec:scheduler}).

\begin{figure}[ht]
\centerline{\includegraphics[width=1\linewidth]{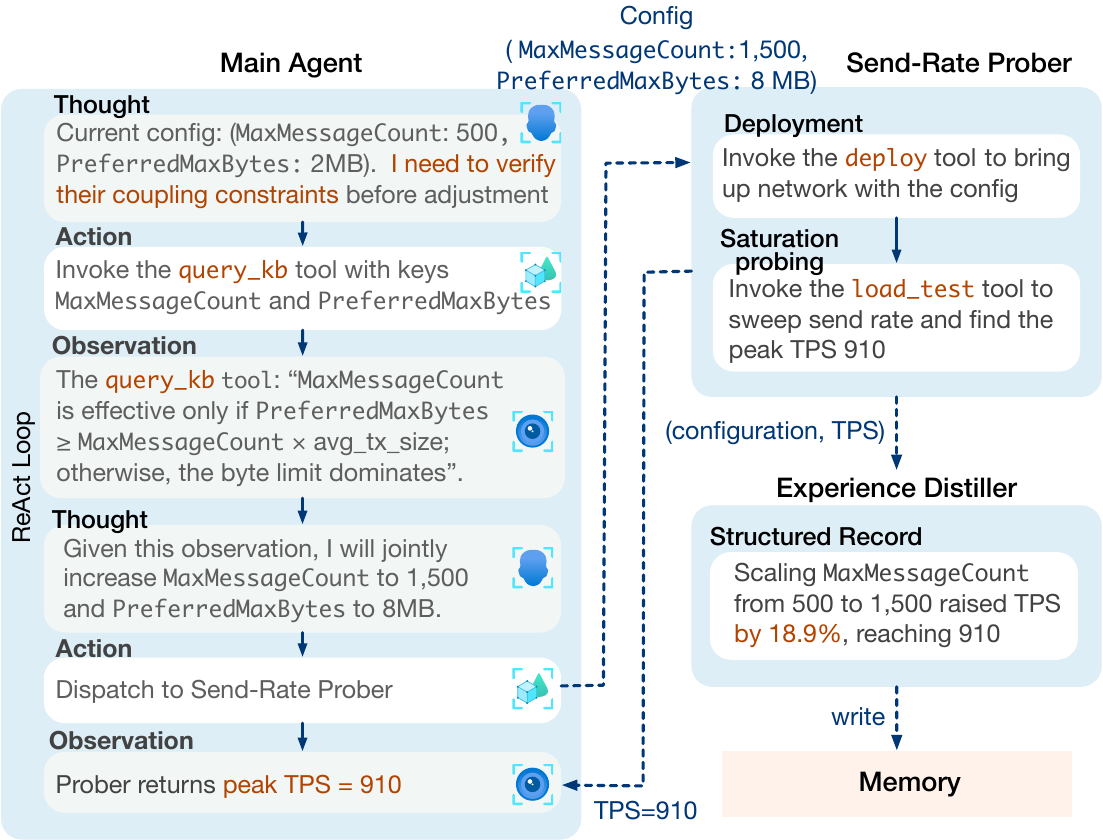}}
\caption{Orchestration Example}
\label{fig:orchestration}
\end{figure}

Fig.~\ref{fig:orchestration} shows an example of the orchestration workflow. The \emph{Main Agent} first queries the knowledge base and learns that ``{\small\texttt{MaxMessageCount}} is effective only if {\small\texttt{PreferredMaxBytes}} $\geq$ {\small\texttt{MaxMessageCount}} $\times$ avg\_tx\_size; otherwise, the byte limit dominates''. It then adjusts {\small\texttt{MaxMessageCount}} to 1,500 and {\small\texttt{PreferredMaxBytes}} to 8\,MB, and dispatches the configuration to the \emph{Send-rate Prober}. The Prober runs deployment and saturation tests, confirming a peak throughput of 910 TPS. The \emph{Experience Distiller} then records this performance delta, closing the loop of decision-making, probing, and knowledge accumulation. Two further mechanisms sustain this workflow: \emph{Tool Mediation}, which governs agent--tool interaction, and \emph{Memory and Context Management}, which maintains agent state and historical experience.

\subsubsection{Tool Mediation}
Every agent capability is exposed through a well-defined tool interface, and every tool invocation is mediated by the Harness, which enforces an access policy: before each agent invocation, only the tool subset assigned to that agent is injected into its execution context.

This access policy yields two benefits. First, it is an action-space pruning mechanism: the \emph{Main Agent} cannot hallucinate low-level operations that are not exposed to it. Second, it provides operational resiliency: when the network degrades severely under load, the \emph{Send-rate Prober} autonomously calibrates the $\mathtt{send\_rate}$ and absorbs the fluctuation, while the Harness keeps raw error states out of the \emph{Main Agent}'s context.

\subsubsection{Memory and Context Management}

\tool{} organizes its memory into three complementary types: \emph{semantic}, \emph{episodic}, and \emph{procedural}~\cite{coala}. Managed by the Harness rather than individual agents, these stores externalize knowledge and experience so that information persists across agent invocations and long-running tuning sessions.

\begin{itemize}
\item \emph{Semantic memory} stores declarative knowledge through two read-only knowledge bases: one describes parameter semantics and coupling dependencies, and the other provides topology-specific probing heuristics. Both are accessed on demand via $\mathtt{query\_kb}$.

\item \emph{Episodic memory} records the tuning trajectory by appending a structured delta record to the experience store after each trial. Raw benchmark logs are stored separately and accessed only through $\mathtt{reflect}$ for historical summarization and $\mathtt{diagnose}$ for performance analysis, keeping them out of the active context.

\item \emph{Procedural memory} maintains the transient runtime state necessary for advancing the optimization loop, including session invariants (e.g., baseline configuration), search progress (e.g., best result), and control signals (e.g., termination status). Unlike the read-only semantic memory and append-only episodic memory, procedural memory is continuously updated.
\end{itemize}

The Harness is the read/write interface to all three stores: agent-facing tools are Harness-mediated access points rather than direct file handles, and the Harness appends the \emph{Experience Distiller}'s delta records to the experience store under the append-only discipline. Procedural memory is exclusively owned by the Harness, so no agent can bypass the memory model or corrupt persistent state.

The context window is the main constraint on an LLM agent~\cite{memgpt}. \tool{} preserves token capacity through a two-pronged strategy. First, a recency-biased summarization mechanism monitors the message list and, past a configurable threshold, compresses distant turns into a structured summary while retaining recent rounds verbatim. Second, long-lived artifacts and large outputs are externalized to persistent storage and accessed only through retrieval tools, so only compact representations remain in the active message list. 

%% file: sections/05_scheduler.tex
\makeatletter
\providecommand{\ext@lstlisting}{lol}
\providecommand{\listingcaption}[1]{\def\@captype{lstlisting}\def\lst@@caption{#1}\caption{#1}}
\makeatother

\section{LLM-Based Predictive Transaction Scheduling}
\label{sec:scheduler}

\begin{figure}[ht]
\centering
\includegraphics[width=\linewidth]{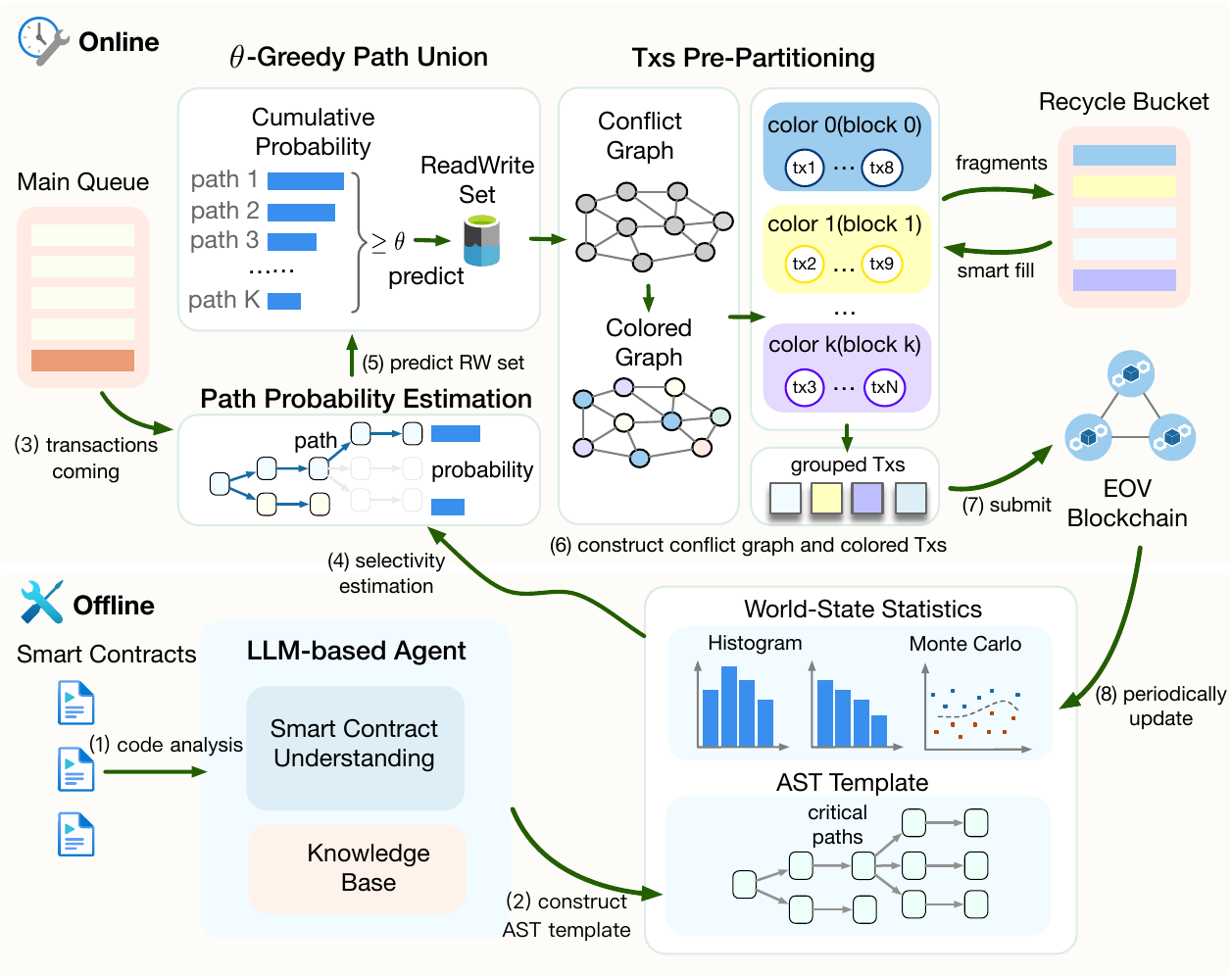}
\caption{Workflow of the Predictive Transaction Scheduler}
\label{fig:scheduler}
\end{figure}

Transaction scheduling relies on a two-phase predictive pipeline to reduce intra-block conflicts, as shown in Fig.~\ref{fig:scheduler}. The offline phase (steps (1)--(2)) analyzes smart contracts to extract AST templates, along with the world-state statistics needed for runtime inference. The online phase (steps (3)--(7)) uses the AST templates and cached state summaries to organize incoming transactions into conflict-free groups, avoiding preventable MVCC conflicts, while the statistics are periodically refreshed from the live world state (step (8)). This allows semantic analysis to be performed once and reused across multiple transactions.

\subsection{Offline Phase: Semantic-Aware Symbolic Execution}
\label{sec:offline}

To predict a transaction's read-write set without executing it, the scheduler must determine the possible execution paths and the keys each path accesses. Conventional \emph{static analysis} of smart contract code~\cite{Bartoletti20} can identify constant key prefixes and simple linear structures, but it provides no basis for estimating which path is likely to be taken for a given transaction and world-state snapshot, and thus cannot distinguish likely access patterns from merely feasible ones. We overcome this limitation by combining \emph{symbolic execution} with \emph{world-state statistical modeling}: the smart contract's state-accessing paths are encoded symbolically offline, and the probability of each path is predicted at scheduling time from statistical summaries of the world state. This shifts the prediction from a static worst-case guess to a context-aware, probabilistic inference of the actual read-write set.

\subsubsection{Critical-Path Extraction and AST Template Generation}
\label{sec:ast_extraction}

As shown in steps (1)--(2) of Fig.~\ref{fig:scheduler}, we combine symbolic path analysis with an LLM-based semantic abstraction. Smart contracts contain many auxiliary paths, such as parameter validation, error handling, and logging, that do not alter the world state but dominate the control-flow graph; in Listing~\ref{lst:tx-savings}(a), for example, the grey branches (lines 2, 4, and 10) do not change which keys are accessed. Enumerating such paths exhaustively is wasteful and obscures the conflict-relevant ones. To ground the LLM agent's code understanding in the domain of smart contracts, we provide it with a curated knowledge base (Fig.~\ref{fig:scheduler}, bottom left) that encodes common smart contract structures and business-logic patterns. Given the source code and the retrieved context, the agent discards auxiliary paths and framework-specific control logic, and retains only the \emph{critical paths} that may influence the resulting read-write set (step (1)). For each retained path it generates a symbolic AST template (step (2)) whose leaves are symbolic \textsc{Read} and \textsc{Write} nodes. Branch conditions along a path are classified into three types, which determine how each branch probability is estimated online:
\begin{itemize}
\item $\mathtt{predicate}$ conditions: depend solely on transaction arguments and can be evaluated immediately online;

\item $\mathtt{linear}$ conditions: compare a single state variable against a constant or argument (e.g., $\mathtt{balance}\ge\mathtt{amount}$);

\item $\mathtt{complex}$ conditions: involve non-linear expressions or multiple state variables.
\end{itemize}

\subsubsection{World-State Statistics Specification}
\label{sec:ws_stats}

Alongside the templates, the LLM agent \emph{declares the set of world-state fields} for which statistical summaries must be maintained. The condition type determines the form of each summary. For $\mathtt{linear}$ conditions, the framework maintains a one-dimensional histogram over the corresponding state field to enable efficient selectivity estimation. For $\mathtt{complex}$ conditions involving multiple correlated variables, it employs \emph{offline Monte Carlo simulation} to pre-compute joint multi-dimensional distributions. Online, the scheduler queries these pre-computed distributions directly, achieving near-instantaneous selectivity estimation and keeping heavy sampling computation off the transaction critical path. To track shifting workload characteristics, both the histograms and the joint distributions are periodically refreshed in the background from the live world state (step (8)). The offline phase thus produces not only symbolic AST templates but also a specification of the world-state statistics required for runtime path-probability estimation, bridging static program structure and dynamic state distributions.

\textit{Example.} Listing~\ref{lst:tx-savings}(b) shows the extracted template, where $\mathtt{arg.0}$ and $\mathtt{arg.1}$ denote the amount and the account identifier. Only two state-altering paths remain: a direct write when $\mathtt{arg.0}\ge 0$ (line 12) and a guarded write when $\mathtt{arg.0}<0$ and $\mathtt{SavingsBalance}\ge -\mathtt{arg.0}$ (lines 6--11). Because the latter carries a $\mathtt{linear}$ condition, the LLM declares $\mathtt{SavingsBalance}$ as a required histogram field (line 1).

\subsection{Online Phase: Read-Write Set Prediction and Pre-Partitioning}
\label{sec:online}

\begin{figure}[!t]
\begin{lstlisting}[language=Go, basicstyle=\ttfamily\scriptsize,
  moredelim={[is][\color{gray}]{^}{^}},
  title={(a) Smart Contract (auxiliary branches in grey)}]
func TransactSavings(stub, args) pb.Response {
  ^if len(args) != 2 { return errormsg(ERR_WRONG_ARGS) }^
  account, err := loadAccount(stub, args[1])
  ^if err != nil { return errormsg(ERR_NOT_FOUND) }^
  amount, _ := strconv.Atoi(args[0])
  if amount < 0 && account.SavingsBalance < -amount {
    return errormsg("Insufficient funds") }
  account.SavingsBalance += amount
  err = saveAccount(stub, account)
  ^if err != nil { return systemerror(err.Error()) }^
  return shim.Success(nil)
}
\end{lstlisting}
\vspace{-1mm}
\begin{lstlisting}[language=json, basicstyle=\ttfamily\scriptsize,
  escapeinside={(*@}{@*)},
  title={(b) Extracted AST template}]
{"(*@\color{blue}\textbf{histograms}@*)": ["(*@\color{blue}\textbf{SavingsBalance}@*)"],
 "functions": {"transact_savings": {"body": [
 {"type": "(*@\color{blue}\textbf{branch}@*)",
  "condition": {"kind": "predicate", "param": "arg.0",
                "op": "<", "rhs": "0"},
  "then": [{"type": "(*@\color{blue}\textbf{branch}@*)",
    "condition": {"kind": "(*@\color{blue}\textbf{linear}@*)",
      "field": "(*@\color{blue}\textbf{SavingsBalance}@*)", "op": "<",
      "rhs": "-arg.0"},
    "then": [],
    "else": [{"type": "(*@\color{blue}\textbf{write}@*)", "key": "arg.1"}]}],
  "else": [{"type": "(*@\color{blue}\textbf{write}@*)", "key": "arg.1"}]}]}}}
\end{lstlisting}
\vspace{-1mm}
\listingcaption{Smallbank \texttt{TransactSavings}: (a) Source Code and (b) the extracted AST template.}
\label{lst:tx-savings}
\end{figure}

\subsubsection{Selectivity Estimation for Path Probabilities}
\label{sec:path_estimation}

At runtime, inbound transactions are buffered in the main queue (step (3) in Fig.~\ref{fig:scheduler}), and the scheduler estimates the traversal probability of each branch node in the pre-extracted AST templates, converting symbolic paths into concrete read-write set predictions. This problem is analogous to \emph{selectivity estimation} in relational query optimizers~\cite{selectivity}. To avoid the prohibitive latency of executing live state queries for every incoming transaction, \tool{} decouples state inspection from scheduling by querying the cached world-state statistics produced offline (step (4)).

For $\mathtt{linear}$ conditions over a single world-state variable, the scheduler uses lightweight, equal-width histograms, cached as JSON files and exposed for $O(1)$ lookups. Branch probabilities are computed as cumulative bin-count ratios over these intervals: for the balance check in Listing~\ref{lst:tx-savings}(a) (line 6), the probability is the fraction of sampled entries in bins satisfying the threshold, with the complement assigned to the insufficient-funds path.

For $\mathtt{complex}$ conditions, the independence assumption behind composing one-dimensional histograms breaks down. The scheduler instead evaluates the condition against the joint distribution pre-computed offline by Monte Carlo simulation (Section~\ref{sec:ws_stats}) and uses the resulting satisfaction ratio as the branch probability.

\subsubsection{$\theta$-Greedy Path Union}
\label{sec:greedy_union}

Once branch traversal probabilities are predicted, the scheduler enumerates all valid execution paths, sorts them in descending order of their joint probabilities, and applies a greedy prefix strategy (step (5) in Fig.~\ref{fig:scheduler}) to construct the minimal subset $\Omega$ such that the cumulative probability reaches or exceeds a confidence threshold $\theta$. The predicted read-write set is the union of world-state keys accessed across all paths in $\Omega$. Sorting descending minimizes $|\Omega|$. This is a minimum-cardinality covering problem~\cite{kellerer2004knapsack}: each path incurs a uniform selection cost (over-prediction risk) but contributes different probability mass. Greedy selection of the most probable paths satisfies the threshold with the fewest paths, which is critical because a larger $\Omega$ expands the union and increases conflict-graph density.

This union-based aggregation introduces an asymmetric safety profile. False positives (predicting a conflict that does not occur) are performance-safe: they merely relegate compatible transactions into separate groups, slightly underutilizing parallelism. False negatives (missing a real conflict) are unsafe: they allow conflicting transactions into the same block, triggering MVCC aborts that waste endorsement and consensus resources. Given this asymmetry, we adopt a near-unity threshold $\theta$, which nearly eliminates false negatives: the predicted read-write set covers the true accessed keys with high confidence, minimizing MVCC aborts. The resulting increase in false positives is harmless, only moderately curtailing parallel throughput without compromising correctness. The scheduler thus systematically errs on the side of over-prediction, trading a small execution capacity for robust conflict avoidance.

\subsubsection{Conflict-Free Pre-Partitioning}
\label{sec:graph_construct}

For each scheduling window of $|V|$ transactions drawn from the main queue, the runtime constructs an undirected conflict graph $G = (V, E)$ (step (6) in Fig.~\ref{fig:scheduler}). Each vertex $v \in V$ encapsulates an incoming transaction accompanied by its predicted read-write set $(R_v, W_v)$ from Section~\ref{sec:greedy_union}. An edge $(u, v) \in E$ is instantiated if and only if the two transactions access at least one common world-state key in a conflicting pattern under MVCC read-conflict semantics~\cite{fabricpp, chacko_failures}. This pairwise intersection check incurs $O(|V|^2)$ time in the worst case.

Once $G$ is constructed, the transaction clustering task can be formally modeled as a variant of the \textit{Bin Packing with Conflicts (BPC)} problem, where all transactions possess equal weight, and the bin capacity is bounded by the target block size $\mathcal{B}$ (corresponding to the {\small\texttt{MaxMessageCount}} parameter). This problem is proven to be NP-hard in the strong sense, and is NP-complete for any fixed capacity $\mathcal{B} \geq 3$. To achieve low scheduling latency at runtime, we eschew exact solvers and adopt the efficient heuristic method (H2)~\cite{bpc2004}. Using an internal priority queue to track vertex degrees during color assignment, this phase runs in $O((|V| + |E|) \log |V|)$ time. BPC tasks are executed across multiple concurrent threads so that scheduling keeps pace with the downstream execution and validation throughput (Section~\ref{sec:eval_scheduling}).

The H2 heuristic yields a set of disjoint color groups $\mathcal{C}$, each a candidate block (color~$k$ / block~$k$ in Fig.~\ref{fig:scheduler}), where each group $G_i \in \mathcal{C}$ is internally conflict-free and bounded by $|G_i| \leq \mathcal{B}$. However, due to uneven workload distributions within the window, some groups may be under-filled; submitting such sparse blocks would waste block space and degrade effective throughput. \tool{} therefore employs a two-level buffer, the main queue and the Recycle Bucket, to shape the sequence before transmission, applying a binary triage to each group $G_i \in \mathcal{C}$:
\begin{itemize}
    \item \textbf{Full Block Group ($|G_i| = \mathcal{B}$):} A group that reaches the full capacity $\mathcal{B}$ is a complete, conflict-free batch. The runtime immediately invokes $\mathtt{ScheduleTransmission}(G_i)$, submitting the grouped transactions to the Fabric network (step (7)).
    \item \textbf{Gapped Block Group ($|G_i| < \mathcal{B}$):} For under-filled groups, the controller attempts a speculative padding routine (\emph{smart fill} in Fig.~\ref{fig:scheduler}). It scans the asynchronous $\mathtt{RecycleBucket}$ (which stores leftover transactions deferred from previous rounds) to extract compatible, conflict-free transactions to fill the gap. If the group absorbs enough transactions to satisfy $|G_i| = \mathcal{B}$, it is immediately emitted via $\mathtt{ScheduleTransmission}(G_i)$. Crucially, to prevent conflict re-introduction from partial fills, if the $\mathtt{RecycleBucket}$ cannot provide sufficient compatible transactions to reach full capacity, the padding operation is aborted, and the entire group $G_i$ is relegated to the $\mathtt{RecycleBucket}$ as a \emph{fragment} to be regrouped in later windows.
\end{itemize}

%% file: sections/06_evaluation.tex
\section{Evaluation}
\label{sec:evaluation}

\subsection{Experimental Setup}

We implement \tool{} on Hyperledger Fabric, a widely adopted EOV permissioned blockchain whose complex configuration space and MVCC-induced transaction conflicts motivate our optimization approach.

\noindent \textbf{Hardware and Topologies.} We evaluate across five Fabric network topologies, ranging from 8 to 40 peers, deployed on public cloud VMs (8 vCPUs, 16 GiB RAM). One VM hosts every four nodes. All five topologies share the same ordering service, consisting of three orderer nodes running the Raft consensus protocol. The endorsement policy is set to MAJORITY in all topologies, i.e., a transaction must be endorsed by peers from a majority of the participating organizations. Each topology uses a dedicated master node running the tuning agent and configuration server, with worker nodes hosting the Fabric peers and orderers. 
 
\noindent \textbf{Implementation.} \tool{} is implemented in Python~3.9.5. All the agents are built on the LangGraph state-graph framework, with DeepSeek-V4-Pro as the underlying reasoning model. Experiments are conducted on Hyperledger Fabric~3.1.3.  

\noindent \textbf{Benchmarks.} We evaluate on three representative smart contracts: \textit{Smallbank}~\cite{fabric_arch} (banking transfers), \textit{Token-ERC-20}~\cite{fabric_arch} (token mint, transfer, and balance queries), and \textit{IOHeavy}~\cite{blockbench} (batch key-value read/write operations). 

\begin{figure*}[t]
  \centering
  \includegraphics[width=\textwidth]{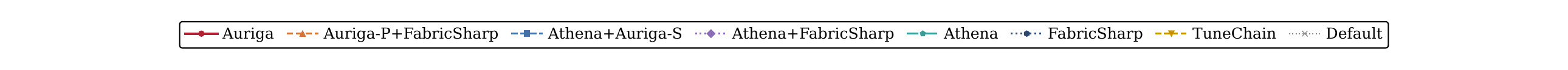}\\[2pt]
  \includegraphics[width=0.95\textwidth]{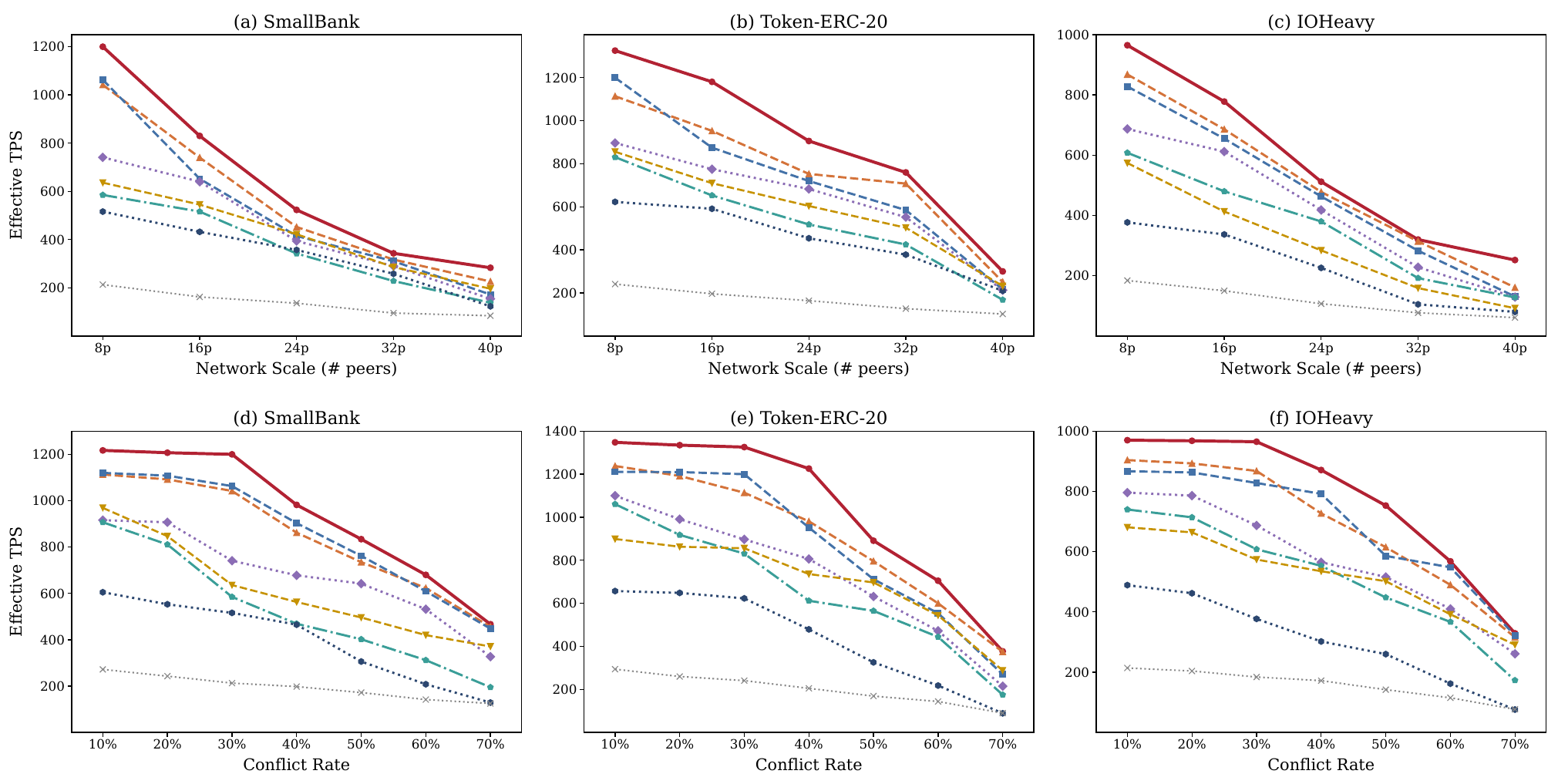}
  \caption{Comparison of effective TPS across different methods: (a--c) performance under varying network scales with a fixed moderate-conflict workload (30\%), and (d--f) performance under varying workload conflict rates on an 8-peer Fabric network.}
  \label{fig:main}
\end{figure*}

\noindent \textbf{Baselines.} Table~\ref{tab:baselines} lists all the compared methods, grouped following the categories of Table~\ref{tab:related-work}. In the end-to-end evaluation (Secs.~\ref{sec:eval_topo} and~\ref{sec:eval_conflict}), we compare \tool{} against the Fabric default configuration, the standalone configuration tuner Athena, the standalone conflict-control system FabricSharp, and TuneChain, the prior approach that jointly targets both factors of Eq.~\eqref{eq:efftps}. We further construct three composite hybrids to isolate and verify the individual performance gains contributed by \emph{our tuning agent and predictive scheduler} when coupled with two representative baselines (Athena and FabricSharp). In the subsystem-level evaluations, \tool-P is additionally compared with the LLM-based tuners BCTuner, AgentTune, and GPTuner (Sec.~\ref{sec:eval_tuning}), and \tool-S with the re-execution-based HTFabric (Sec.~\ref{sec:eval_scheduling}). For a fair comparison, every method that involves configuration tuning (\tool{}, \tool-P, Athena, TuneChain, BCTuner, AgentTune, GPTuner, and the hybrids built on \tool-P or Athena) is given the same budget of 50 tuning rounds.

\begin{table}[!ht]
\centering
\footnotesize
\caption{Compared methods}
\label{tab:baselines}
\begin{tabularx}{\linewidth}{@{}l >{\raggedright\arraybackslash}X@{}}
\toprule
\textbf{Method} & \textbf{Description} \\
\midrule
Default & Fabric default configuration, no scheduling \\
\addlinespace
\multicolumn{2}{@{}l}{\textit{Configuration tuning}} \\
Athena~\cite{athena} & Multi-agent DDPG for configuration tuning, no scheduling \\
BCTuner~\cite{deng2026bctuner} & LLM-guided Monte Carlo tree search over blockchain configurations \\
AgentTune~\cite{agenttune} & LLM-guided tree-beam search \\
GPTuner~\cite{gptuner} & LLM-guided Bayesian optimization \\
\addlinespace
\multicolumn{2}{@{}l}{\textit{Conflict control}} \\
FabricSharp~\cite{fabricsharp} & Transaction reordering via pre-filter, no configuration tuning \\
HTFabric~\cite{song2024htfabric} & WARD block reordering with parallel re-execution of invalid transactions \\
HTFabric-WARD & HTFabric with WARD reordering only (no re-execution) \\
\addlinespace
\multicolumn{2}{@{}l}{\textit{Joint optimization}} \\
TuneChain~\cite{tunechain} & PPO-based online tuning of block size and block interval, with read-only-first block reordering toggled as a tuning knob \\
\addlinespace
\multicolumn{2}{@{}l}{\textit{Ours and composite hybrids}} \\
\tool{}-P & \tool{}'s configuration tuning (Sec.~\ref{sec:tuning})\\
\tool{}-S & \tool{}'s scheduling (Sec.~\ref{sec:scheduler})\\
\tool{}  & Our complete method \\
Athena+\tool-S & Athena-tuned configuration combined with \tool{}'s scheduler \\
\tool-P+FabricSharp & \tool{}'s optimal configuration applied to the FabricSharp network \\
Athena+FabricSharp & Athena-tuned configuration on a FabricSharp network \\
\bottomrule
\end{tabularx}
\end{table}

\subsection{Cross-Topology Performance}
\label{sec:eval_topo}

As shown in Fig.~\ref{fig:main}(a--c), \tool{} consistently achieves the highest effective TPS across all topologies and benchmarks. On the 8-peer Smallbank benchmark, \tool{} reaches 1{,}200 eff-TPS, outperforming TuneChain by 89\% and Athena by 105\%. Augmenting existing SOTA baselines with \tool{}'s individual components substantially enhances their performance, but the composite hybrids still fall short of our complete system. The largest margin over the best-performing standalone baseline, $1.97\times$, occurs on IOHeavy at 40 peers (252 vs.\ 128 eff-TPS for Athena).

While absolute performance degrades as the topology grows from 8 to 40 peers, \tool{} maintains its advantage over all competitors across all three benchmarks, validating its scalability at larger topologies. On Smallbank, \tool{} at 40 peers (283 eff-TPS) outperforms TuneChain (196, $+$44\%), Athena (138, $+$105\%), and FabricSharp (123, $+$130\%). TuneChain remains the strongest standalone baseline on Smallbank and Token-ERC-20 at every scale, while Athena is the strongest on IOHeavy. On Smallbank and Token-ERC-20, TuneChain's edge over Athena widens with network size (from $+$9\% to $+$42\% and from $+$3\% to $+$38\% between 8 and 40 peers, respectively).
This follows from its online block-size/block-interval action space, whose optimum is scale-sensitive (Finding~1, Sec.~\ref{sec:eval_tuning}). Nevertheless, \tool{} outperforms TuneChain on all three benchmarks. TuneChain's limitations explain this gap: it tunes only restart-free knobs, leaving gossip, communication, and concurrency parameters at defaults, and its reordering only moves read-only transactions ahead of writes within a block, so it cannot resolve conflicts among write transactions.

\subsection{Cross Contention Performance}
\label{sec:eval_conflict}

Fig.~\ref{fig:main}(d--f) shows system behavior as the conflict rate scales from 10\% to 70\%. Following the workload-generation protocol of~\cite{fabricsharp}, we calibrate the hot-key fraction via simulation to match each target conflict rate under the operational block size. \tool{} achieves the highest eff-TPS at every conflict rate on all three benchmarks, with the largest margins under moderate contention. At 70\% conflict, \tool{}-P+FabricSharp nearly closes the gap (e.g., 375 vs.\ 377 eff-TPS on Token-ERC-20), as FabricSharp's pre-filter retains a small, nearly conflict-free subset, and \tool{}-P's configuration amplifies its throughput (Sec.~\ref{sec:eval_tuning}).

Standalone baselines fail complementarily as contention rises. Athena's success rate collapses from $\sim$90\% at 10\% conflict to 29--43\% at 70\%, because it lacks conflict control, though its tuned capacity keeps eff-TPS above standalone FabricSharp. FabricSharp preserves near-perfect validation success only by discarding conflicting transactions before ordering; dropped transactions fail silently and are counted as failures. TuneChain degrades more gracefully than Athena: from 10\% to 70\% conflict, its eff-TPS drops by 57--68\% versus 77--84\% for Athena, as its reordering activates when invalid transactions surge; at 70\%, TuneChain leads Athena on all three benchmarks (e.g., 371 vs.\ 195 eff-TPS on Smallbank). However, the reordering is reactive and type-level, and cannot prevent conflicts among write transactions. \tool{} leads TuneChain at every conflict rate, with the margin peaking at moderate contention (30--40\%), where key-level prediction still separates nearly all conflicting transactions into different blocks.

\subsection{Configuration Tuning Evaluation}
\label{sec:eval_tuning}

To isolate \tool-P's performance, we evaluate it under conflict-free workloads to measure configuration quality, with all tuning methods running for 50 tuning rounds per topology.

\begin{figure}[htbp]
  \centering
  \vspace{-3pt}
  \includegraphics[width=\linewidth]{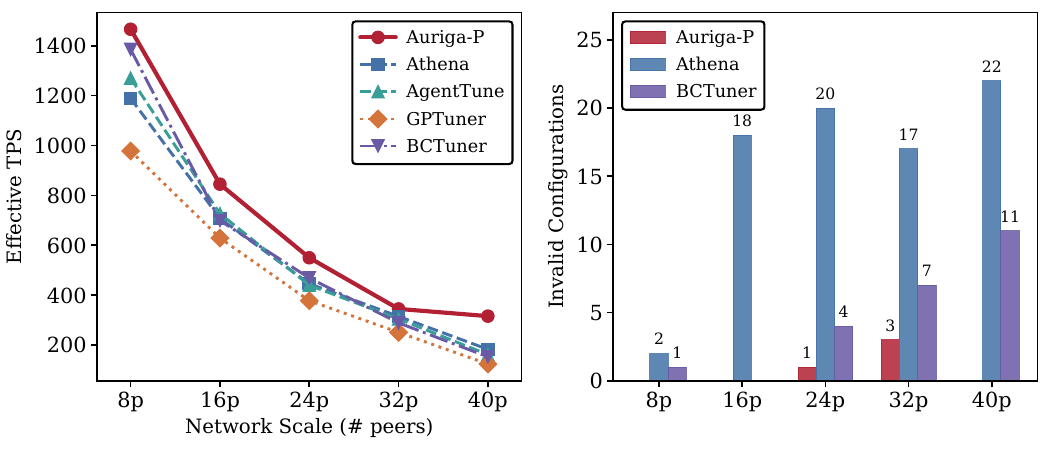}
  \vspace{-5pt}
  \caption{Configuration tuning evaluation: (a) effective TPS across five Fabric network topologies, and (b) non-viable configurations proposed during tuning by \tool-P, Athena, and BCTuner.}
  \label{fig:param_tuning}
\end{figure}

As shown in Fig.~\ref{fig:param_tuning}(a), \tool-P achieves the highest effective TPS across all topologies. BCTuner, the most recent LLM-guided blockchain tuner, is the strongest baseline at 8, 24, and 32 peers, confirming the value of blockchain-specific LLM grounding; yet \tool-P outperforms it by 6--21\%. \tool-P's advantage widens with scale: at 40 peers, it reaches 316 eff-TPS, exceeding Athena, BCTuner, AgentTune, and GPTuner by 73\%, 77\%, 95\%, and 155\%, respectively. BCTuner's gap is largest at 40 peers, where it even falls slightly behind Athena (179 vs.\ 183 eff-TPS), because fixed-send-rate scoring misranks candidates once the sustainable send rate drops with topology size. \tool-P instead drives each candidate to saturation via the \emph{Send-rate Prober}, ranking configurations by actual capacity regardless of scale.

Fig.~\ref{fig:param_tuning}(b) shows that \tool-P remains stable across network scales, proposing near-zero non-viable configurations (those triggering cluster collapse or consensus deadlocks). In contrast, Athena is more susceptible to knob constraints, frequently generating configurations that destabilize the network. BCTuner also yields more non-viable configurations than \tool-P, rising to 11 at 40 peers, for the same fixed-send-rate reason.

\noindent \textbf{Tuning Overhead.} We profile \tool-P during a representative 50-step tuning run on the 8-peer topology (Table~\ref{tab:tuning_cost}). The run completes in about 3.90\,h (281\,s per step), yet LLM inference accounts for only about 24\% of the wall-clock time (the remainder is deployment and benchmark execution), and 79.85\% of prompt tokens hit DeepSeek's prompt cache, substantially reducing inference cost. The agent process itself remains lightweight, blocked on I/O for over 90\% of the run with mean CPU utilization of 0.11\%.

\begin{table}[t]
\centering
\small
\caption{Resource consumption of \tool-P over a 50-step tuning run on the 8-peer topology}
\label{tab:tuning_cost}
\footnotesize
\begin{tabular}{l r}
\toprule
\textbf{Metric} & \textbf{Value} \\
\midrule
\multicolumn{2}{l}{\textit{Tuning run}} \\
\quad Wall-clock time & 14{,}048\,s ($\approx$3.90\,h) \\
\quad Optimization steps & 50 \\
\quad LLM calls (total / per step) & 292 / 5.84 \\
\quad Avg.\ latency per LLM call & 11.58\,s \\
\midrule
\multicolumn{2}{l}{\textit{Token usage per step}} \\
\quad Prompt tokens & $\approx$48.8k \\
\quad Completion tokens & $\approx$4.7k \\
\quad Prompt cache hit ratio & 79.85\% \\
\midrule
\multicolumn{2}{l}{\textit{Agent process (sampled at 1\,Hz)}} \\
\quad CPU util.\ (mean / p95 / max) & 0.11\% / 1.0\% / 28.0\% \\
\quad Memory util.\ (mean / p95 / max) & 0.77\% / 0.83\% / 0.94\% \\
\bottomrule
\end{tabular}
\end{table}

\noindent \textbf{Insights from the Tuning Process.} \tool-P's exploration traces reveal several structural observations about Fabric's parameter landscape. 

\begin{figure}[htbp]
  \centering
  \includegraphics[width=\linewidth]{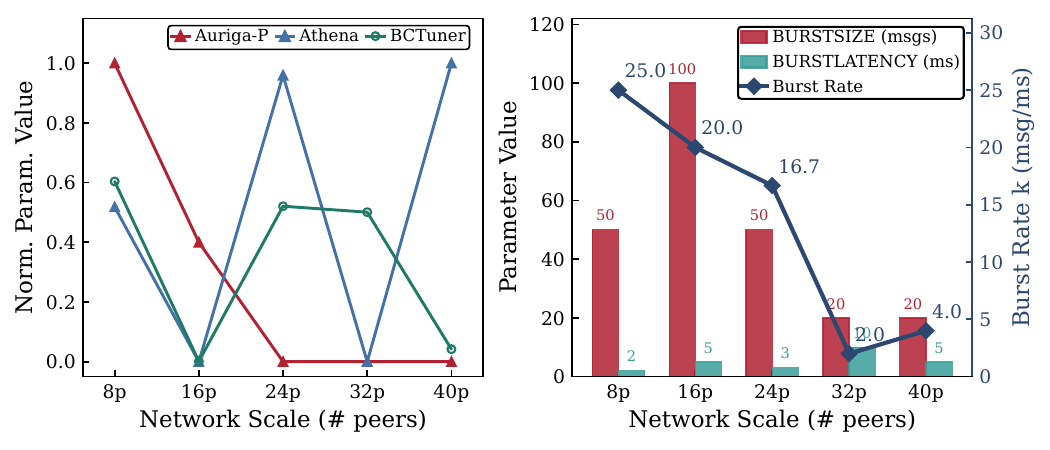}
  \caption{Parameter exploration insights: (a) block-cutting adaptations, and (b) coupled gossip burst parameter dynamics.}
  \label{fig:combined_insights}
\end{figure}

\begin{figure*}[t]
  \centering
  \includegraphics[width=0.7\textwidth]{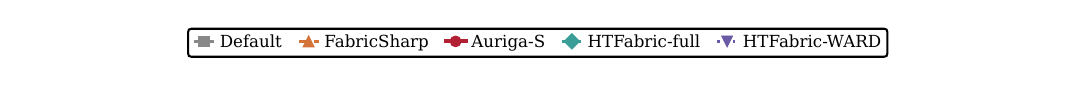}\\ 
  \includegraphics[width=0.85\textwidth]{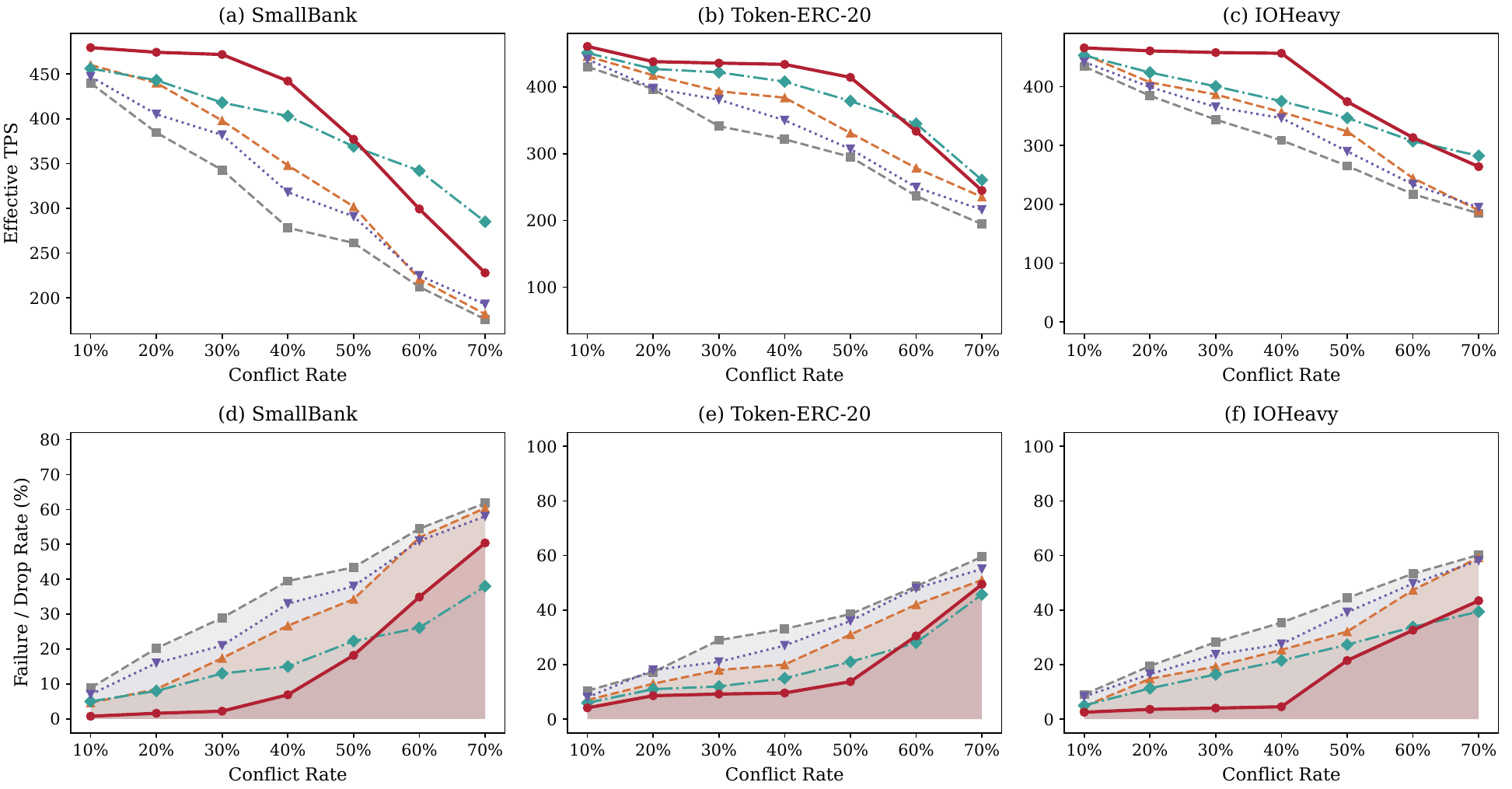}
  \caption{Transaction scheduling comparison. Top: effective TPS; bottom: failure rate, with FabricSharp pre-filter drops counted as failures.}
  \label{fig:scheduling}
\end{figure*}

 \textbf{Finding 1: Inverse Scaling of Block-Cutting Parameters.} As the network expands from 8 to 40 peers, \tool-P dynamically shrinks block-cutting parameters ($\mathtt{MaxMessageCount}$ and $\mathtt{PreferredMaxBytes}$), with $\mathtt{MaxMessageCount}$ decreasing by $25\times$ (Fig.~\ref{fig:combined_insights}a). This shift reflects a trade-off: larger blocks amortize fixed overhead in small networks, whereas smaller blocks mitigate straggler latency across endorsement chains. \tool-P captures this scale-dependent relationship; Athena appears less sensitive to network scale.

 \textbf{Finding 2: Latent Parameter Coupling Discovery.} \tool-P uncovers hidden parameter interdependencies overlooked by existing literature, such as the gossip layer's $\mathtt{MAXPROPAGATIONBURSTSIZE}$ and $\mathtt{MAXPROPAGATIONBURSTLATENCY}$. These knobs jointly dictate the burst rate as $\text{SIZE}/\text{LATENCY}$. As shown in Fig.~\ref{fig:combined_insights}(b), \tool-P scales this rate from 25~msg/ms at 8 peers down to 2~msg/ms at 32 peers to adapt to topology-driven gossip pressure. In contrast, the baselines tend to optimize these parameters independently.

\subsection{Transaction Scheduling Evaluation}
\label{sec:eval_scheduling}
Throughout this subsection, all experiments are conducted on an 8-peer topology with all block-cutting parameters fixed to the FabricSharp~\cite{fabricsharp} reference value (\texttt{MaxMessageCount}\,=\,200), isolating conflict resolution. We compare the arrival-order \textbf{Default} baseline, \textbf{FabricSharp}'s orderer-side filtering, \textbf{HTFabric}'s block reordering with parallel re-execution, and \textbf{\tool-S}'s conflict-free grouping, plus \textbf{HTFabric-WARD}, which applies only WARD reordering to separate HTFabric's two mechanisms. The evaluation spans the workloads of Section~\ref{sec:eval_conflict} with conflict rates swept from 10\% to 70\%. All methods are driven at a fixed send rate of 500 tx/s.

Fig.~\ref{fig:scheduling} shows effective throughput across all scenarios. Compared with the unscheduled \textbf{Default} baseline, \tool-S improves performance on all benchmarks and at all conflict rates by 7--59\% by eliminating most MVCC aborts, and it outperforms FabricSharp in every setting by up to 39\%. FabricSharp's failure rate climbs from 4.6--7.0\% at 10\% conflict to 51.0--60.4\% at 70\% (dropped transactions count as failures), whereas \tool-S keeps its failure rate below 10\% up to 40\% conflict. Compared with the re-execution alternative HTFabric, \tool-S leads under low-to-moderate contention (10--50\%) on all benchmarks (by up to 22\% on IOHeavy at 40\%). HTFabric-WARD alone trails \tool-S in every setting, so HTFabric's gains come from re-execution. This gain comes with a trade-off: transactions invalidated by MVCC conflicts are re-executed by a single fast peer after ordering, so their final execution results are no longer endorsed by the peers designated in the endorsement policy~\cite{huang2026}. This relaxes the endorsement guarantee on which Fabric's trust model relies, a concern for deployments with strict security or compliance requirements. HTFabric overtakes \tool-S only under extreme contention (60--70\%, except IOHeavy at 60\%) by 3--25\%, but requires kernel modification (WARD in the ordering service, re-execution on a fast peer), whereas \tool-S operates entirely at the client gateway; the two are complementary, with pre-execution avoidance removing predictable conflicts and re-execution salvaging residual ones where kernel changes are acceptable.

\textbf{Read-Write Set Prediction Accuracy.} Since false negatives can place conflicting transactions in the same block, we measure \tool-S's prediction accuracy ($\theta=0.95$) on 30{,}000 transactions (10{,}000 per benchmark). Coverage exceeds 99.6\% on all benchmarks and reaches 99.86\% overall, and only 86 of 6{,}533 MVCC aborts (1.3\%) are attributable to prediction error (Table~\ref{tab:rw_accuracy}), showing that the near-unity threshold suppresses false negatives.

\begin{table}[t]
\centering
\small
\caption{Read-write set prediction accuracy of \tool-S ($\theta = 0.95$, 10{,}000 transactions per benchmark)}
\label{tab:rw_accuracy}
\footnotesize
\setlength{\tabcolsep}{3pt}
\begin{tabular}{l r r r r r}
\toprule
\textbf{Benchmark} & \textbf{Coverage} & \textbf{Key-FN} & \textbf{MVCC} & \textbf{Attrib.} & \textbf{Ratio} \\
\midrule
Smallbank    & 99.98\% & 0.02\% & 944     & 6  & 0.6\% \\
Token-ERC-20 & 99.62\% & 0.38\% & 3{,}183 & 70 & 2.2\% \\
IOHeavy      & 99.98\% & 0.02\% & 2{,}406 & 10 & 0.4\% \\
\midrule
\textbf{Total} & \textbf{99.86\%} & \textbf{0.14\%} & \textbf{6{,}533} & \textbf{86} & \textbf{1.3\%} \\
\bottomrule
\end{tabular}
\vspace{2pt}

\parbox{\linewidth}{\scriptsize MVCC: number of MVCC aborts; Attrib.: MVCC aborts attributable to read-write set misprediction; Ratio: Attrib./MVCC.}
\end{table}

\begin{figure}[t]
  \centering
  \includegraphics[width=\linewidth]{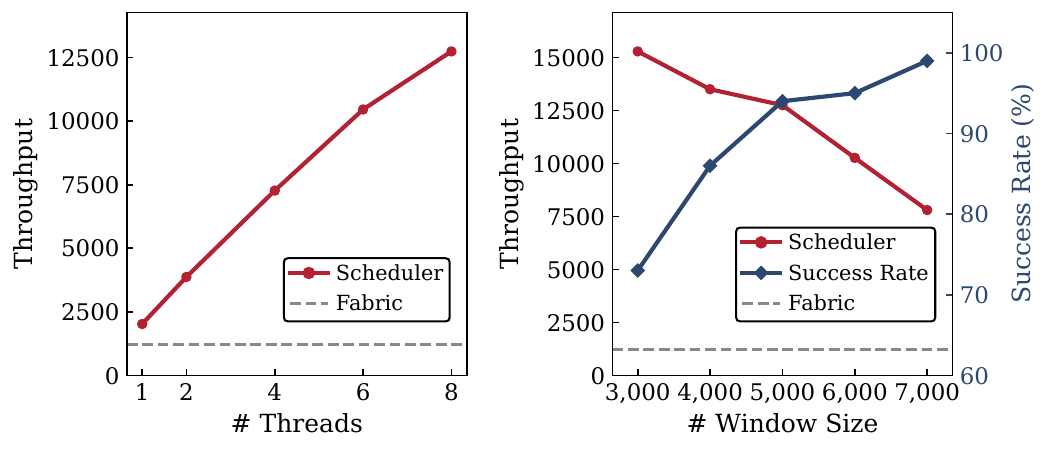}
  \caption{(a) throughput scaling with scheduler thread count, and (b) throughput and success rate as a function of scheduling window size (Smallbank).}
  \label{fig:scheduler_overhead}
\end{figure}

\textbf{Scheduling Overhead and Thread Scaling.} To verify that the scheduler's $\mathcal{O}(V^2)$ graph construction does not bottleneck the pipeline, we benchmark it on Smallbank against a reference Fabric commit ceiling of about 1,200 tx/s, the peak effective throughput observed on this benchmark (Fig.~\ref{fig:main}): throughput scales from 2,021 tx/s with one thread to 12,742 tx/s with 8 threads ($6.3\times$, Fig.~\ref{fig:scheduler_overhead}(a)), and even single-threaded execution clears the ceiling by $1.7\times$, leaving $10.6\times$ headroom with the default 8 threads.

\textbf{Window Size Optimization.} Sweeping the scheduling window from 3,000 to 7,000 transactions (Fig.~\ref{fig:scheduler_overhead}(b)) reveals a trade-off: larger windows lift the success rate from 73\% to 99\% but reduce throughput through denser conflict graphs. A performance knee occurs at 5,000, where expanding from 4,000 gains 8 percentage points of success rate (86\% to 94\%) at only a 6\% throughput cost; we adopt 5,000 as the default.

%% file: sections/08_conclusion.tex
\section{Conclusion and Future Work}
\label{sec:conclusion}

This paper presents \tool{}, a performance optimization system that jointly manages configuration tuning and transaction scheduling in permissioned blockchains. By combining harness-governed LLM agents for safe parameter exploration with a predictive scheduler for proactive concurrency control, \tool{} reduces MVCC conflicts prior to execution. Evaluations show that \tool{} improves effective throughput by up to $6.0\times$ over the default Fabric configuration and $2.0\times$ over SOTA baselines. For future work, we plan to extend this dual-optimization framework to multi-chain and cross-chain environments, and investigate adaptive scheduling mechanisms.

%% file: sections/09_author_biographies.tex
\vspace{2\baselineskip}
\begingroup
\setlength{\topsep}{0pt}\setlength{\partopsep}{0pt}
\setlength{\itemsep}{0pt}\setlength{\parsep}{0pt}\setlength{\parskip}{0pt}
\footnotesize

\newcommand{\mybio}[3]{%
  \par\noindent
  \begin{minipage}[c]{0.82in}%
    \includegraphics[width=0.7in,height=0.875in,clip,keepaspectratio]{#1}%
  \end{minipage}%
  \hspace{4pt}%
  \begin{minipage}[c]{\dimexpr\linewidth-0.90in\relax}%
    \textbf{#2}\ #3%
  \end{minipage}%
  \par\vspace{15pt}%
}

\mybio{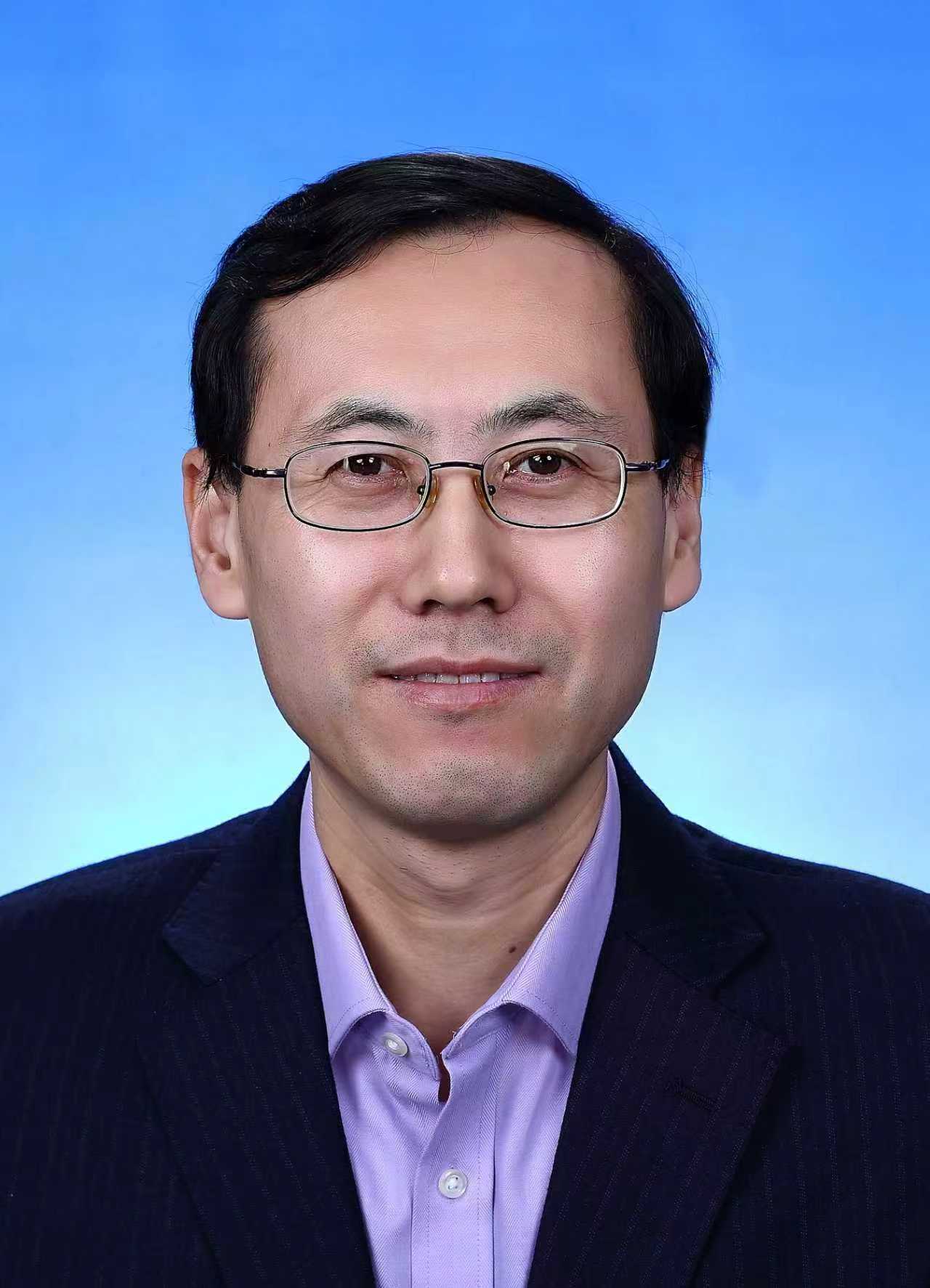}{Xiao Zhang}
{received the PhD degree from the Institute of Computing Technology, CAS. He is a professor at Renmin University of China. His research interests include big data systems, database architecture, blockchain, and benchmarking.}

\mybio{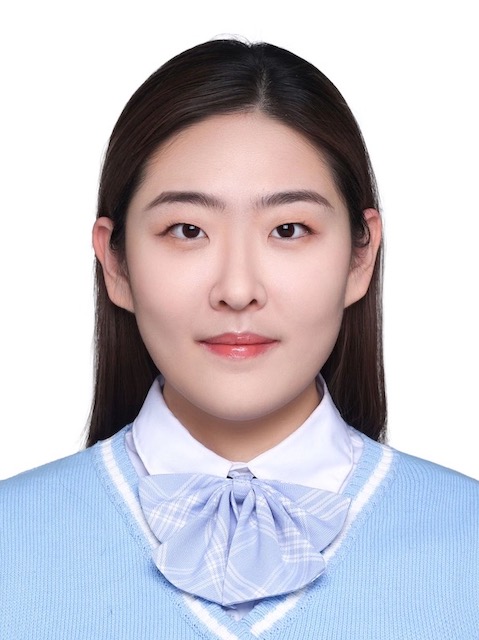}{Tianyang Wang}
{received the BE degree from Renmin University of China and MEng degree from Nanyang Technological University, Singapore. She is pursuing the PhD degree at Renmin University of China. Her research focuses on blockchain systems and performance optimization.}

\mybio{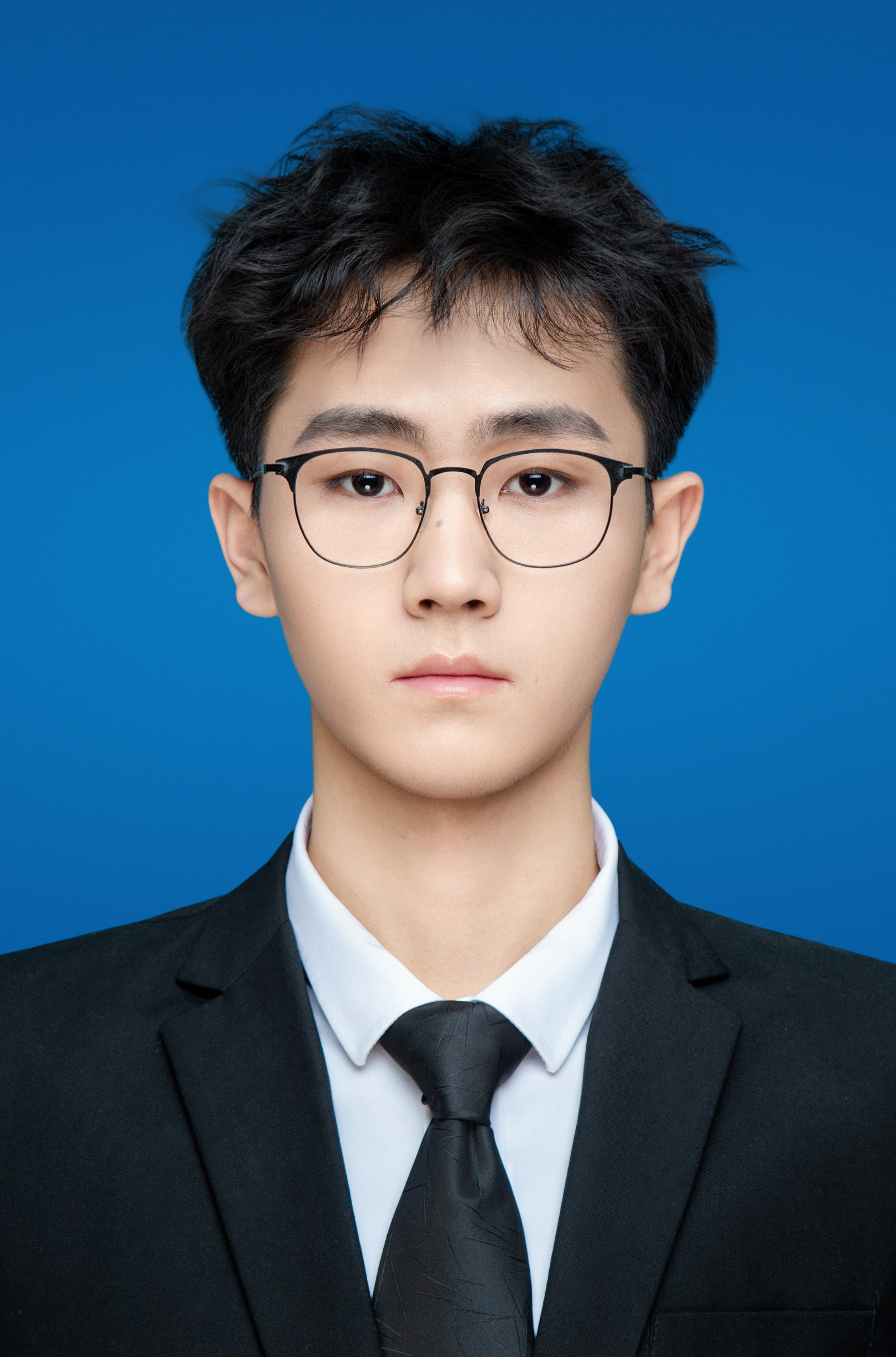}{Fangtao Gu}
{received the BE degree from Renmin University of China, where he is pursuing the MS degree. His research focuses on LLM agents.}

\mybio{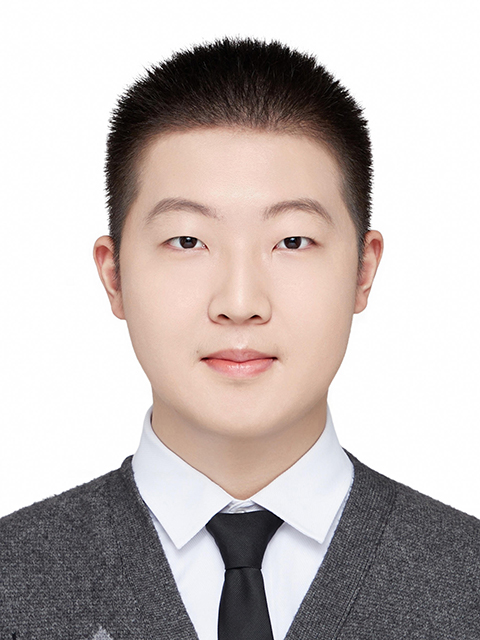}{Haoxiang Sun}
{is pursuing the BS degree at Renmin University of China. His research interests include LLM reasoning and reinforcement learning.}

\mybio{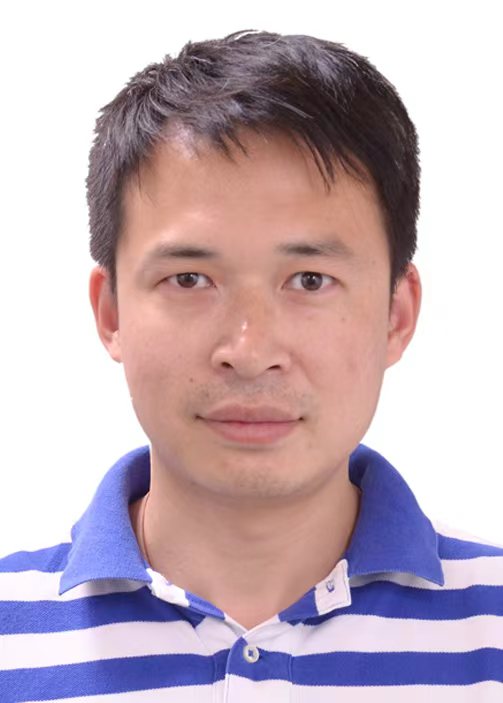}{Puwei Wang}
{received the PhD degree from the Institute of Computing Technology, CAS. He is an associate professor at Renmin University of China. He has published 30+ papers and received ICWS 2023 Best Paper and DIKW 2021 Outstanding Paper Awards. His research interests include LLM-based agents, blockchain, and services computing.}

\mybio{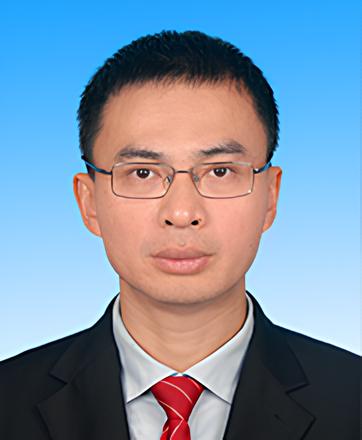}{Jinchuan Chen}
{received the PhD degree from The Hong Kong Polytechnic University. He is a professor at Renmin University of China. He has published 30+ papers in VLDB, ICDE, VLDBJ and TSC etc. His research focuses on distributed data management and blockchain.}

\endgroup